\documentclass[pre,reprint,twocolumn,nofootinbib,groupedaddress,floatfix]{revtex4-2}

\usepackage{amsmath}
\usepackage{amsfonts}
\usepackage{amssymb}
\usepackage{mathtools}
\usepackage{empheq}
\usepackage{multirow}

\usepackage[pdftex]{graphicx}
\usepackage[usenames,dvipsnames,pdftex]{xcolor}

\usepackage[varg]{txfonts}
\usepackage{newtxtext}
\AtBeginDocument{
\fontsize{10.5pt}{\baselineskip}\selectfont
}
\usepackage{titlesec}
\titleformat*{\section}{\centering\fontsize{10.5pt}{\baselineskip}\selectfont\bfseries}
\titleformat*{\subsection}{\centering\fontsize{10.5pt}{\baselineskip}\selectfont\bfseries}
\titleformat*{\subsubsection}{\centering\fontsize{10.5pt}{\baselineskip}\selectfont\itshape}

\usepackage{bm}
\usepackage{upgreek}

\usepackage{physics}
\usepackage{mathrsfs}

\hypersetup{
    colorlinks=true,
    citecolor=blue,
    linkcolor=blue,
    urlcolor=blue,
}

\newcommand{\X}{\mathrm{X}}
\newcommand{\eq}{\mathrm{eq}}
\newcommand{\tot}{\mathrm{tot}}
\newcommand{\st}{\mathrm{st}}
\newcommand{\sol}{\mathrm{sol}}
\newcommand{\E}{\mathrm{E}}
\newcommand{\ES}{\mathrm{ES}}
\renewcommand{\P}{\mathrm{P}}
\newcommand{\Q}{\mathrm{Q}}
\renewcommand{\S}{\mathrm{S}}
\newcommand{\refr}{\mathrm{ref}}
\newcommand{\cl}{\mathrm{cl}}
\renewcommand{\op}{\mathrm{op}}
\DeclareMathOperator*{\argmin}{argmin}
\DeclareMathOperator*{\argmax}{argmax}
\newcommand{\tildebm}[1]{\tilde{\bm{#1}}}

\newcommand{\ueta}{\ensuremath{\upeta}}
\newcommand{\utheta}{\ensuremath{\uptheta}}
\DeclareMathAlphabet{\mathsfit}{T1}{\sfdefault}{\mddefault}{\sldefault}
\newcommand{\res}{\mathrm{res}}
\newcommand{\chg}{\mathrm{ch}}
\newcommand{\cyc}{\mathrm{cy}}

\allowdisplaybreaks

\def\SetMake#1|#2?{\left\{#1\,\middle|\,#2\right\}}
\newcommand{\Set}[1]{{\SetMake#1?}}
\def\DMake#1||#2?{D\left(#1\,\middle\|\,#2\right)}
\newcommand{\D}[1]{{\DMake#1?}}

\makeatletter
\def\hush#1{
\def\hushm@de{b}
{\@tfor\member:=#1\do{%
\if\hushm@de b \member\def\hushm@de{f}
\else\hspace{-0.1em}\member
\fi}}}
\makeatother

\makeatletter
\def\Tns#1{%
\def\tens@rmode{n}
{\@tfor\member:=#1\do{%
\if^\member\def\tens@rmode{p}
\else\if_\member\def\tens@rmode{b}
\else
\if\tens@rmode n \member\fi
\if\tens@rmode b {{}_{\member}}\fi
\if\tens@rmode p {{}^{\member}}\fi
\def\tens@rmode{n}
\fi\fi}}
}
\makeatother

\begin{document}

\title{\texorpdfstring{Information-geometric structure for chemical thermodynamics:\\
An explicit construction of dual affine coordinates}{Information-geometric structure for chemical thermodynamics: An explicit construction of dual affine coordinates}}
\date{\today}

\author{Naruo Ohga}
\author{Sosuke Ito}

\affiliation{Department of Physics, Graduate School of Science, The University of Tokyo, \\7-3-1 Hongo, Bunkyo-ku, Tokyo 113-0033, Japan}

\begin{abstract}
We construct an information-geometric structure for chemical thermodynamics, applicable to a wide range of chemical reaction systems including non-ideal and open systems. For this purpose, we explicitly construct \textit{dual affine coordinate systems}, which completely designate an information-geometric structure, using the extent of reactions and the affinities of reactions as coordinates on a linearly-constrained space of amounts of substances. The resulting structure induces a metric and a divergence (a function of two distributions of amounts), both expressed with chemical potentials. These quantities have been partially known for ideal-dilute solutions, but their extensions for non-ideal solutions and the complete underlying structure are novel. The constructed geometry is a generalization of dual affine coordinates for stochastic thermodynamics. For example, the metric and the divergence are generalizations of the Fisher information and the Kullback--Leibler divergence. As an application, we identify the chemical-thermodynamic analog of the Hatano--Sasa excess entropy production using our divergence.
\end{abstract}
 
\maketitle

\section{Introduction}

Chemical thermodynamics, despite its long history~\cite{gibbs1878equilibrium,dedonder1936thermodynamic,prigogine1967introduction,beard2008chemical, kondepudi2014modern}, has recently significantly advanced through an analogy with stochastic thermodynamics~\cite{rao2016nonequilibrium,ge2016nonequilibrium}. Chemical thermodynamics is formulated on deterministic rate equations~\cite{feinberg2019foundations}, and therefore its physical nature is quite different from stochastic thermodynamics~\cite{schnakenberg1976network,seifert2008stochastic,sekimoto2010stochastic,seifert2012stochastic}. Nevertheless, their mathematical structures are similar when we identify probability in stochastic thermodynamics with the amounts of substances in chemical thermodynamics. Through this analogy, several existing results and concepts in stochastic thermodynamics have been imported and further developed in chemical thermodynamics ~\cite{polettini2014irreversible,ge2016mesoscopic,ge2017mathematical,falasco2018information,wachtel2018thermodynamically,avanzini2019jcp,penocchio2019thermodynamic,falasco2019negative,avanzini2020thermodynamics,yoshimura2021information,avanzini2021nonequilibrium,yoshimura2021prl,avanzini2022thermodynamics}.

One of the theoretical tools in stochastic thermodynamics is information geometry, useful for deriving universal inequalities and finding a decomposition of quantities, two major tasks in this field. Information geometry is a differential geometry on the space of probability distributions~\cite{amari2007methods,amari2016information}. It provides a unified geometric structure over informational quantities~\cite{cover2006elements}, such as the Fisher information~\cite{fisher1922mathematical} and the Kullback--Leibler (KL) divergence~\cite{kullback1951information}, and offers further geometrical insights, such as the shortest path and the projection~\cite{amari2016information}. Initially developed in the realm of information theory and statistics~\cite{chentsov1982statistical,amari1982differential,amari1995information,amari2001information}, information geometry has been imported to statistical physics~\cite{tanaka2000information,crooks2007measuring,edward2008length,brody2008jpa,sivak2012prl,polettini2013nonconvexity,machta2015dissipation,aguilera2021natcommun} and stochastic thermodynamics~\cite{ito2018prl,gupta2020tighter,nicholson2020time,ito2020prx,Bryant2020PNAS,zhang2020information,ito2022information,nakamura2019reconsideration,shiraishi2019prl,kolchinsky2021pre,kolchinsky2021prx,ohgaito2021,ito2020unified,kolchinsky2021dependence} thanks to the probabilistic nature of these fields. For example, the information-geometric metric and path length has been utilized for thermodynamic speed limits, uncertainty relations, and other bounds on dissipation~\cite{ito2018prl,gupta2020tighter,nicholson2020time,ito2020prx,zhang2020information,Bryant2020PNAS,ito2022information}. Geometric decomposition of the KL divergence has led to decompositions and lower bounds of entropy productions~\cite{shiraishi2019prl,nakamura2019reconsideration,ohgaito2021,kolchinsky2021pre,kolchinsky2021prx}. The connection between the fluctuation theorems~\cite{seifert2012stochastic} and information geometry has yielded other decompositions of entropy production~\cite{ito2020unified,kolchinsky2021dependence}.

Information geometry has recently been linked to chemical thermodynamics, but the link remains weak and insufficient for a broad application to finding universal inequalities and decomposing quantities in chemical thermodynamics. For example, the generalized Fisher information was used in Ref.~\cite{yoshimura2021information} to derive a geometrical speed limit for ideal dilute solutions. The KL divergence for non-normalized distributions~\cite{csiszar1991least} has also been used to rewrite the Gibbs free energy of ideal dilute solutions~\cite{shear1967analog, higgins1968some,horn1972general,rao2016nonequilibrium, ge2016nonequilibrium, ge2016mesoscopic, yoshimura2021information}. However, these connections are limited to ideal dilute solutions. Moreover, they are based on the apparent similarities between \textit{quantities} in information geometry and chemical thermodynamics, not rooted in information-geometric \textit{structures}. If we can find an information-geometric structure for chemical thermodynamics, we can expect to obtain geometric insights into chemical thermodynamics that helps to derive physical results systematically. We can also expect that the structure will allow a natural generalization to non-ideal solutions~\cite{avanzini2021nonequilibrium,ge2017mathematical}. The major difficulty in finding such a structural link is that chemical thermodynamics does not have a probabilistic nature.

Information geometry is rooted in a general mathematical framework called \textit{dually flat geometry}~\cite{amari2007methods,amari2016information}, which is not limited to the space of probability distributions. Dually flat geometry specifies a geometric structure by a pair of two coordinate systems called \textit{dual affine coordinates}, which contain the complete information to calculate all the geometrical quantities. Various dual affine coordinates have been constructed in dually flat geometry. For example, the dual affine coordinates for conventional information geometry on the space of probability distributions have long been known~\cite{amari1985differential}. In our previous paper~\cite{ohgaito2021}, we found another construction of dual affine coordinates for information geometry that uses the stochastic-thermodynamic total entropy. This construction would be a suitable reference to construct dual affine coordinates for chemical thermodynamics.

In this paper, we construct dual affine coordinates for chemical thermodynamics using the Gibbs free energy and the chemical potentials. The construction is applicable even to non-ideal solutions. We also generalize the construction for open chemical reaction systems, i.e., those exchanging some molecules with the surroundings, by replacing the Gibbs free energy with the cumulative entropy production. The main consequences of the construction are as follows: (i) It provides a complete information-geometric structure for chemical thermodynamics. (ii) In particular, we obtain a divergence (a function of two distributions of amounts of substances), expressed using the chemical potential difference and the amounts of substances. We also obtain an information-geometric metric, which is the Hessian of the Gibbs free energy. If the system is an ideal dilute solution, these quantities reduce to the KL divergence and the Fisher information for non-normalized distributions. (iii) As an application, we consider open systems admitting a nonequilibrium steady state. We find that the divergence between any state and the steady state serves as an effective potential function that takes the minimum at the steady state. This effective potential identifies a `closed counterpart' of an open system with the same geometric structure.

This paper is organized as follows. In Secs.~\ref{sec_chemical} and \ref{sec_geometry}, we present the preliminaries on chemical thermodynamics and dually flat geometry, respectively. In Sec.~\ref{sec_closed}, the main section of this paper, we construct dual affine coordinates for closed chemical reaction systems and discuss the consequences.  Section~\ref{sec_open} generalizes the construction to open chemical reaction systems. In Sec.~\ref{sec_steadystate}, we apply the geometry to find an effective potential function in open systems.

\section{Chemical Thermodynamics}
\label{sec_chemical}

\begin{figure}[b]
    \centering
    \includegraphics[width=\hsize,clip]{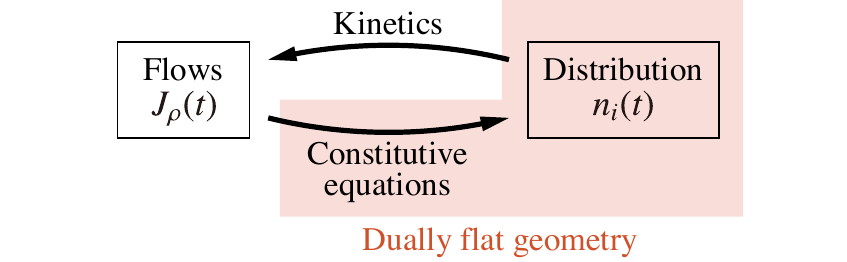}
    \caption{The theoretical structure of chemical thermodynamics. Our dually flat geometry for chemical thermodynamics relies on the distribution and the constitutive equation.}
    \label{fig_flow_distribution}
\end{figure}

We consider two classes of chemical reaction systems: closed systems and open systems. A closed system does not exchange particles with the surroundings, while an open system does. In both systems, the time evolution is described by two types of variables: \textit{distributions} and \textit{flows}. The interplay between these two variables is determined by \textit{kinetics} and \textit{constitutive equations} (Fig.~\ref{fig_flow_distribution}). We will introduce these concepts one by one, together with associated thermodynamic quantities. Our dually flat geometry of chemical thermodynamics is based only on distributions and constitutive equations.

\begin{figure}
    \centering
    \includegraphics[width=\hsize,clip,page=2]{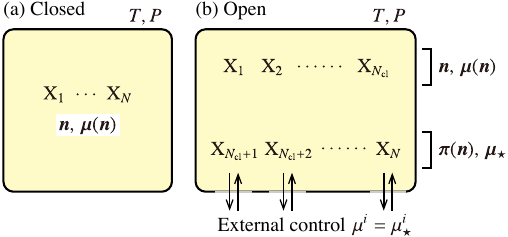} \caption{(a) A closed system with chemical species $\X_1,\dots,\X_N$. The amounts of substances are $\bm n$, and the chemical potentials are $\bm \mu(\bm n)$. (b) An open system. The closed species are denoted by $\X_1,\dots,\X_{N_\cl}$. The corresponding amounts of substances are $\bm n$, and the chemical potentials are $\bm \mu(\bm n)$. The open species are denoted by $\X_{N_\cl+1},\dots,\X_N$. The corresponding amounts of substances are $\bm\pi(\bm n)$, and the chemical potentials are kept constant at $\bm\mu_\star$. The details of the control scheme are irrelevant to our construction.}
    \label{fig_systems}
\end{figure}

\subsection{Closed systems} 

\subsubsection{System} 
\label{subsec_ch_system}

A closed system is a mixture of $N$ chemical species $\X_1,\dots,\X_N$ (including solvent species) that do not exchange particles with the surroundings [Fig.~\ref{fig_systems}(a)]. The chemical species undergo $M$ chemical reactions
\begin{align}
    \Tns{\lambda_1^\rho} \X_1 + \dots + \Tns{\lambda_N^\rho} \X_N
    \,\rightleftarrows\,
    \Tns{\kappa_1^\rho} \X_1 + \dots + \Tns{\kappa_N^\rho} \X_N,
    \label{ch_reaction}
\end{align}
for $\rho=1,\dots,M$. Here, $\Tns{\lambda_i^\rho}$ and $\Tns{\kappa_i^\rho}$ are nonnegative constants called stoichiometric coefficients\footnote{We distinguish super- and subscripts according to the convention in differential geometry to reflect the geometric structure of our results.}. We define the \textit{stoichiometric matrix} by $\mathsfit S \equiv (\Tns{S_i^\rho} ){}_{i=1}^N {}_{\rho=1}^M$ with $\Tns{S_i^\rho} \coloneqq \Tns{\kappa_i^\rho} - \Tns{\lambda_i^\rho}$, which quantifies the change of $\X_i$ by the $\rho$th reaction. We assume that all the reactions are reversible at least at a very low rate. We also assume that the reactions conserve the mass. Therefore, reactions with births or deaths of molecules (such as $0\rightleftarrows \X_i$ or $\X_i \rightleftarrows 0$) are prohibited. 

We always assume the system to be uniform and with no hydrodynamic flows. In other words, the degrees of freedom other than chemical reactions, such as diffusion, mixing, heat conduction, and hydrodynamic motion, are assumed to relax very fast compared to chemical reactions. We also assume that the systems are kept at constant temperature $T$ and pressure $P$. Therefore, the state of the system is completely specified by the amounts of the species.

\subsubsection{Distribution}
\label{subsec_ch_distribution}

One of the variables accounting for the time evolution is the \textit{distribution}, i.e., the set of molar amounts of substances $\bm{n}\equiv (n_i)_{i=1}^N \equiv (n_1,\dots,n_N)^\mathsf{T}$, treated as a column vector. The distribution evolves in time, written as $\bm n(t)$. 

The Gibbs free energy $G(\bm n)$ and the chemical potentials $\mu^i(\bm n)$ are functions of $\bm n$. We suppose the Gibbs free energy is given. We define the chemical potential by $\mu^i(\bm n) \coloneqq \partial G/ \partial n_i$, where the partial derivative is under the constant $T$ and $P$. Therefore, an infinitesimal change in $\bm n$, denoted by $d\bm n$, incurs
\begin{equation}
dG = 
\sum_i \mu^i(\bm{n}) dn_i,
\label{ch_infinitesimal}
\end{equation}
where $dG$ is the infinitesimal change in $G$. We write $\bm\mu \equiv(\mu^1,\dots,\mu^N)$. Hereafter, we always treat a vector indexed with a subscript (such as $\bm n\equiv(n_i)_{i=1}^N$) as a column vector and a vector indexed with a superscript (such as $\bm\mu\equiv(\mu^i)_{i=1}^N$) as a row vector.

Gibbs free energy must satisfy the extensive property
\begin{equation}
    G(\lambda\bm n) = \lambda G(\bm n)
\end{equation}
for an arbitrary constant $\lambda > 0$. The extensive property leads to the Euler relations~\cite{fermi1956thermodynamics,kondepudi2014modern}
\begin{equation}
G(\bm{n}) = 
\sum_i \mu^i(\bm{n}) n_i,
\label{ch_Euler}
\end{equation}
which will be used repeatedly in this paper.

Gibbs free energy must also satisfy the convexity~\cite{tasaki2022}
\begin{equation}
    G(\lambda \bm n + (1-\lambda ) \bm n') \leq \lambda G(\bm n) + (1-\lambda) G(\bm n')
    \label{ch_convexity}
\end{equation}
for any two amounts of substances $\bm n,\bm n'$ and any $0 <  \lambda < 1$. The convexity is not strict, i.e., the inequality $\leq$ in Eq.~\eqref{ch_convexity} cannot be replaced by $<$ for two reasons. First, the extensive property implies that the inequality is saturated if $\bm n \propto \bm n'$. Second, if phase separations or phase transitions occur, the inequality can be saturated even if $\bm n $ is not proportional to $\bm n'$. For simplicity, we exclude the second situation by assuming the absence of phase separations or phase transitions, although we can easily relax this assumption. For mathematical simplicity, we also assume that the Gibbs free energy is twice-differentiable.

\subsubsection{Flows}
\label{subsec_ch_flows}

The other variable is the \textit{flow}, i.e., the rate of reactions at time $t$, denoted by $J_\rho(t)$. More precisely, the rate $J_\rho(t)$ is one mole per unit time when the $\rho$th reaction transforms $\Tns{\lambda_i^\rho}$ moles of $\X_i$ into $\Tns{\kappa_i^\rho}$ moles of $\X_i$ per unit time.

Associated with the flows is the entropy production rate. Defining the affinity~\cite{gibbs1878equilibrium} (thermodynamic force) of the $\rho$th reaction by
\begin{equation}
    {F}^\rho(\bm n) \coloneqq -\sum_i  \mu^i(\bm n) \Tns{S_i^\rho},
    \label{ch_affinity}
\end{equation} 
the entropy production rate due to the $\rho$th reaction is independent of the other reactions and given by $T^{-1}F^\rho J_\rho$ \cite{kondepudi2014modern,polettini2014irreversible}. Since we have assumed that no degrees of freedom other than chemical reactions are out of equilibrium, the total entropy production rate $\dot\Sigma(t)$ is given by
\begin{equation}
    \dot\Sigma (t)
    = \frac{1}{T} \sum_\rho F^\rho(\bm n(t)) J_\rho(t).
    \label{ch_affinitytosigma}
\end{equation}
A state with vanishing affinity for all the reactions is called an equilibrium state and denoted by $\bm n^\eq$.

\subsubsection{Kinetics}
\label{subsec_ch_kinetics}

The flow $J_\rho(t)$ is determined from the distribution $\bm n(t)$ through \textit{kinetics} of the form 
\begin{equation}
    J_\rho(t) = \mathscr K_\rho(\bm n(t)),
    \label{ch_kinetics}
\end{equation}
where $\mathscr K_\rho$ is a function of $\bm n$. In this paper, we do not use the specific form of kinetics and just regard $\mathscr K_\rho(\bm n)$ as given so that the second law of thermodynamics $\sum_\rho F^\rho(\bm n) \mathscr K_\rho(\bm n)\geq 0$ holds. For example, a class of kinetics consistent with the thermodynamics of an ideal dilute solution is the mass action kinetics~\cite{lund1965guldberg,horn1972general,feinberg2019foundations}.

\subsubsection{Constitutive equations}
\label{subsec_ch_constitutive}

Constitutive equations determine how flows change the distribution in time and, combined with kinetics, completely determine the time evolution. The constitutive equations of closed systems is
\begin{equation}
    \frac{dn_i}{dt} = \sum_\rho \Tns{S_i^\rho} J_\rho(t).
    \label{ch_constitutive}
\end{equation}
Namely, the change in amounts of substances is solely due to chemical reactions. With this constitutive equation, the entropy production is written solely in terms of distribution. Indeed, using Eqs.~\eqref{ch_infinitesimal}, \eqref{ch_affinity}, \eqref{ch_affinitytosigma}, and \eqref{ch_constitutive}, we obtain
\begin{equation}
    \dot\Sigma(t) = -\frac{1}{T} \frac{dG}{dt}.
    \label{ch_sigma_G}
\end{equation}
In other words, $-G(\bm n)/T$ equals the cumulative entropy production up to an additive constant.

\subsection{Open systems}

Open systems inherit most of the definitions and assumptions from closed systems. Here we discuss the differences from closed systems.

\subsubsection{System}

An open system is a mixture of $N$ species that exchanges some species with the surroundings~\cite{polettini2014irreversible} [Fig.~\ref{fig_systems}(b)]. We assume that the first $N_\cl$ species are closed species, i.e., those not exchanged with the surroundings, and the other $(N - {N_\cl})$ species are open species, i.e., those exchanged with the surroundings. Namely, $\X_1,\dots,\X_{N_\cl}$ are closed, and $\X_{N_\cl+1},\dots,\X_N$ are open. 

We restrict ourselves to systems in which open species are \textit{potentiostatted}, i.e., the chemical potentials of the open species are fixed at prescribed values. An open system with potentiostatting is one of the simplest models of chemical reaction systems admitting a nonequilibrium steady state. 

\begin{figure}
    \centering
    \includegraphics[width=\hsize,clip,page=3]{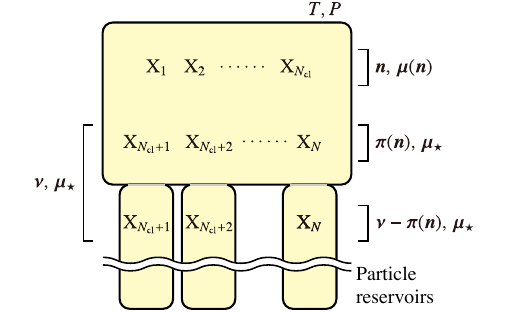} \caption{One specific scheme to realize an open system. The system exchanges the open species with the particle reservoirs. The reservoirs are so large (represented by the wavy line) that their chemical potentials can be considered constant in time. The symbols for the amounts of substances ($\bm n$, $\bm\pi(\bm n)$, and $\bm\nu$) and the chemical potentials ($\bm\mu(\bm n)$ and  $\bm\mu_\star$) are shown (see Secs.~\ref{subsec_ch_opendistribution} and \ref{subsec_ch_reservoir}).}
    \label{fig_systems_reservoir}
\end{figure}

Potentiostatting can be realized by several different schemes, and our results do not rely on the detail of the scheme. For example, it is realized by attaching vast particle reservoirs of the open species to the system through semipermeable membranes that allow only the open species to pass (Fig.~\ref{fig_systems_reservoir}), supposing that the exchanges of the open species are much faster than the chemical reactions. Alternatively, one can precisely control the amounts of open species within the system to keep their chemical potentials constant (See Ref.~\cite{avanzini2022thermodynamics} for a discussion on the equivalence between different control schemes).

We assume all the other assumptions in Sec.~\ref{subsec_ch_system} as they are. In particular, the system involves $M$ reactions as in Eq.~\eqref{ch_reaction}. The $M$ reactions do not include the exchanges of species with the surroundings.

\subsubsection{Distribution}
\label{subsec_ch_opendistribution}

We write the amounts of the closed species by $\bm n \equiv (n_i)_{i=1}^{N_\cl}$ and those of the open species by $\bm\pi \equiv (\pi_i)_{i=N_\cl+1}^N$. We introduce the Gibbs free energy $G(\bm n,\bm\pi)$ and the chemical potentials $\mu^i(\bm n,\bm\pi)$ as in Sec.~\ref{subsec_ch_distribution}, only replacing $\bm n$ with $(\bm n,\bm\pi)$. The chemical potentials of the open species are potentiostatted as
\begin{equation}
    \mu^i(\bm n, \bm \pi) = \mu^i_\star \qquad (\text{$i$: open species}).
    \label{ch_mudesig}
\end{equation} 
We use the abbreviation $\bm\mu(\bm n,\bm \pi) \equiv \qty(\mu^i(\bm n,\bm \pi))_{i=1}^{N_\cl}$ for the chemical potentials of the closed species and $\bm \mu_\star \equiv (\mu^i_\star)_{i=N_\cl+1}^N$ for those of the open species.

The amounts of open species $\bm\pi$ are instantaneously determined by potentiostatting~\eqref{ch_mudesig}. We assume the existence of the $\bm\pi$ that solves \eqref{ch_mudesig} for each $\bm n$. This assumption is physically natural since, if we fix $\bm\mu_\star$ in the protocol in Fig.~\ref{fig_systems_reservoir}, we can expect $\bm\pi$ to settle down at a certain value. Due to the assumption that the Gibbs free energy is strictly convex except for two distributions with $(\bm n,\bm\pi)\propto (\bm n',\bm\pi')$, the solution of Eq.~\eqref{ch_mudesig} is unique for each $\bm n$. The unique solution is a function of $\bm n$ and denoted by $\bm\pi(\bm n)$. Therefore, the independent variables of a potentiostatted open system are $\bm n$. Hereafter, we always implicitly assume $\bm\pi = \bm\pi(\bm n)$ and write $G(\bm n) \equiv G(\bm n,\bm\pi(\bm n))$ and $\mu^i(\bm n) \equiv \mu^i(\bm n,\bm\pi(\bm n))$.

\subsubsection{Flows, kinetics, and constitutive equations}

The flows, the entropy production rate, and the kinetics are introduced in the same way as in Secs.~\ref{subsec_ch_flows} and \ref{subsec_ch_kinetics} with the substitution of $\bm n$ with $(\bm n, \bm\pi)$. We will use the abbreviation $F^\rho(\bm n) \equiv F^\rho(\bm n,\bm\pi(\bm n)) $ and $\mathscr K_\rho(\bm n) \equiv \mathscr K_\rho(\bm n,\bm\pi(\bm n))$.

The only difference from closed systems lies in the constitutive equations. The amounts of the closed species are determined solely by the reactions, while those of open species are by potentiostatting:
\begin{equation}
\begin{alignedat}{2}
    &\frac{dn_i}{dt} = \sum_\rho \Tns{S_i^\rho} J_\rho(t)
    &\qquad& (\text{$i$: closed species}),\\
    &\pi_i = \pi_i(\bm n) 
    &\qquad& (\text{$i$: open species}).
\end{alignedat}
    \label{ch_constitutive_open}
\end{equation}

Open systems may relax to a steady state $\bm n^\st$, i.e., a state with $dn_i/dt = \sum_\rho \Tns{S_i^\rho} \mathscr K_\rho(\bm n^\st) =0$ for $i=1,\dots,N_\cl$.  In general, a steady state is not determined from thermodynamic quantities (such as affinities) but depends on the details of the kinetics. A steady state may be nonequilibrium, i.e., accompanied by a nonzero entropy production rate. Therefore, the cumulative entropy production may not be determined solely by the distribution in open systems.

\subsubsection{A specific realization with particle reservoirs}
\label{subsec_ch_reservoir}

Although our formulation does not depend on the detailed scheme of potentiostatting, we will exploit the specific realization in Fig.~\ref{fig_systems_reservoir} in constructing the geometry. In this realization, the total system, i.e., the principle system and the particle reservoirs together, can be regarded as a closed system.

In this specific realization, the total amount of an open species held in the system and the reservoirs together is well-defined, which we write as $\bm\nu \equiv (\nu_i)_{i=N_\cl+1}^N$. The difference $\bm\nu - \bm\pi$ gives the amounts of open species in the particle reservoirs. The set $(\bm n, \bm\nu) $ constitutes the independent variables of the total system. The total amounts should be vast, and we only need their changes during processes. Nevertheless, we use $\bm\nu$ itself to make the discussions transparent. 

The total Gibbs free energy $G^\tot(\bm n,\bm\nu)$ is the sum of the Gibbs free energy of the system and those of the reservoirs. We approximate $G^\tot$ so that the chemical potentials of the open species are fixed at $\bm\mu_\star$ (see Appendix~\ref{sec_appendix_Gtot}):
\begin{equation}
    G^\tot(\bm n, \bm\nu) = \bm\mu(\bm n) \cdot \bm n 
    + \bm\mu_\star \cdot \bm\nu 
    + \mathrm{const.},
    \label{ch_Gtot_approx}
\end{equation}
where the constant does not depend on $(\bm n,\bm\nu)$. The total Gibbs free energy equals the cumulative entropy production up to an additive constant.

\subsection{Example: Ideal dilute solutions}

An important class of chemical reaction systems is the ideal dilute solutions~\cite{fermi1956thermodynamics, rao2016nonequilibrium,ge2016nonequilibrium}. In this paper, we do not necessarily assume the ideal dilute property, but we use ideal dilute solutions for an explicit example of our results. 

Consider that one of the species $\X_1$ is the solvent species accounting for most of the amount of substances in the system, while the other species $\X_2,\dots,\X_N$ are the small amounts of solutes dissolved in the solvent. We assume $n_i/n_1 \ll 1$ for $i=2,\dots,N$. We further assume the ideal dilute property, i.e., that there is no interaction between solute molecules, and the free energy change due to dissolving the solutes into the solvent is linearly proportional to $n_i$. Under these assumptions, the Gibbs free energy is~\cite{fermi1956thermodynamics,tasaki2022}
\begin{align}
G(\bm n) &= 
\sum_{i\neq1} \qty[ \strut \qty(\strut \mu_\circ^i + RT \ln \frac{n_i}{n_1} ) n_i - RT n_i] + \mu_\circ^1 n_1,
\end{align}
where $R$ is the gas constant, and $\mu_\circ^i$ is a constant depending only on $T$ and $P$. By differentiating the Gibbs free energy by $n_i$, we obtain the chemical potentials
\begin{subequations}
\label{ch_mu_ideal_all}
\begin{align}
    &\mu^1(\bm n) = \mu_\circ^1 - RT \sum_{i\neq 1} \frac{n_i}{n_1} ,
    \label{ch_mu_ideal_solvent}\\
    &\mu^i(\bm n) = \mu_\circ^i + RT\ln \frac{n_i}{n_1} \qquad (i=2,\dots,N).
    \label{ch_mu_ideal_solute}
\end{align}
\end{subequations}
In particular, the chemical potential of a solute species $\X_i$ is solely determined by $n_i$ and $n_1$. 

For open ideal dilute solutions, we always treat the solvent as a closed species. Nevertheless, we need not to completely shut down the exchange of the solvent with the surroundings. We can allow $n_1$ to change in the order of the amounts of solutes $n_i\, (\ll n_1)$. We discuss this point later within the demonstration of our results for ideal dilute solutions.

The amounts of the open solutes $\bm\pi(\bm n)$ is explicitly calculated from Eq.~\eqref{ch_mudesig}:
\begin{equation}
    \pi_i(\bm n) = n_1 \exp( \frac{\mu^i_\star - \mu_\circ^i}{RT})  \qquad (\text{$i$: open species}).
    \label{ch_pi_ideal_solute}
\end{equation}
In particular, if the solvent does not react with the solute species, $n_1$ is kept constant, and therefore $ \pi_i$ is a constant. In this case, fixing the chemical potentials of solutes is equivalent to fixing their amounts.

\section{Dually flat geometry}
\label{sec_geometry}

Dually flat geometry is the mathematical framework underlying information geometry. In this section, we introduce dually flat geometry as a general mathematical framework without being specific to thermodynamics. We will hereafter use the terms `information geometry' and `dually flat geometry' interchangeably.

\subsection{Dual affine coordinates and the quadruplet}

\begin{figure}[!b]
    \centering
    \includegraphics[width=\hsize,clip,page=1]{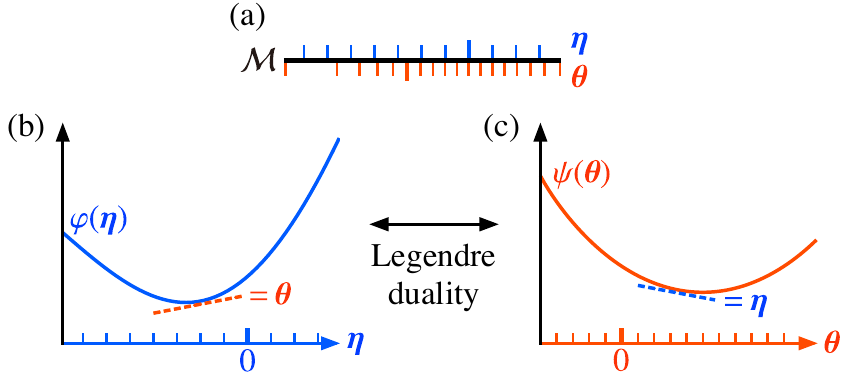}
    \caption{Schematics of a dually flat geometry, sketched for a one-dimensional space (although one dimension is too simple for practical uses). (a) The one-dimensional space $\mathcal M$ (black), equipped with two coordinate systems $\bm\eta$ and $\bm\theta$. (b) The \ueta-coordinates and the convex function $\varphi(\bm\eta)$. The slope of $\varphi(\bm\eta)$ gives the other coordinate $\bm\theta$. (c) The \utheta-coordinates and the convex function $\psi(\bm\theta)$. The slope of $\psi(\bm\theta)$ gives the coordinate $\bm\eta$.}
    \label{fig_duallyflatgeometry1}
\end{figure}

Dually flat geometry is a variant of differential geometry. It introduces two different coordinate systems, $\bm\eta$ and $\bm \theta$, into one space (manifold) [Fig.~\ref{fig_duallyflatgeometry1}(a)] and provides a geometric structure unifying the two coordinates~\cite{amari2007methods,amari2016information}. The two coordinate systems together are called the \textit{dual affine coordinates}.

To construct a dually flat geometry on a space $\mathcal M$, we need to provide a coordinate system $\bm \eta$ on $\mathcal M$ and a twice-differentiable strictly-convex function $\varphi(\bm \eta)$ as inputs. The coordinate system, called the $\eta$-\textit{coordinates}, is written as $\bm\eta \equiv (\eta_1,\dots,\eta_K)^\mathsf{T}$, where $K$ is the dimensionality of the space $\mathcal M$. The range of the \ueta-coordinates is $\mathcal{V}\subseteq\mathbb{R}^K$. The coordinates $\bm\eta\in\mathcal V$ must admit a one-to-one correspondence with the points $\P\in\mathcal M$, denoted by $\bm \eta(\P)$ and $\P(\bm\eta)$. The function $\varphi(\bm\eta)$ must be strictly convex, i.e., its $\hush{K\times K}$ Hessian matrix   $(\partial^2\varphi/\partial\eta_\rho\partial\eta_\sigma)$ must be positive-definite at any $\bm\eta\in\mathcal V$.

We perform the complete Legendre transform of $\varphi(\bm\eta)$ with respect to $\bm\eta$ to obtain the other coordinate system $\bm\theta\equiv(\theta^1,\dots,\theta^K)$ and a new convex function $\psi(\bm\theta)$:
\begin{subequations}
\label{ge_defofthetapsi}
\begin{align}
  \theta^\rho(\bm\eta) &\coloneqq \pdv{\varphi}{\eta_\rho},
  \label{ge_defoftheta} \\
  \psi(\bm\theta) &\coloneqq \sum_{\rho=1}^K \theta^\rho \eta_\rho(\bm\theta) - \varphi(\bm\eta(\bm\theta)) .
  \label{ge_defofpsi}
\end{align}
\end{subequations}
The new coordinates $\bm\theta$ are called the $\theta$-\textit{coordinates}. Its range is naturally determined from the construction \eqref{ge_defoftheta} and denoted by $\mathcal{U} \subseteq\mathbb{R}^K$. Due to the strict convexity of $\varphi(\bm\eta)$, we obtain one-to-one correspondences between $\bm\eta\in\mathcal V$, $\bm\theta\in\mathcal U$ and the points $\P\in \mathcal M$. We denote these correspondences as $\bm\eta(\P)$, $\bm\theta(\P)$, $\P(\bm\eta)$, $\P(\bm\theta)$, $\bm\eta(\bm\theta)$, and $\bm\theta(\bm\eta)$. The new function $\psi(\bm\theta)$ is strictly convex with respect to the \utheta-coordinates. We readily have the inverse Legendre transformation
\begin{subequations}
\label{ge_invetavarphi}
\begin{align}
  \eta_\rho(\bm\theta) &= \pdv{\psi}{\theta^\rho} ,
  \label{ge_inveta}\\
  \varphi(\bm\eta) &= \sum_{\rho=1}^K \eta_\rho \theta^\rho(\bm\eta) - \psi(\bm\theta(\bm\eta)) . \label{ge_invvarphi} 
\end{align}
\end{subequations}

We put together the two coordinate systems and the two convex functions and call them the quadruplet $(\bm\eta,\varphi(\bm\eta),\bm\theta,\psi(\bm\theta))$ [Fig.~\ref{fig_duallyflatgeometry1}(b)(c)]. The quadruplet suffices to designate all the geometric structures of a dually flat geometry, as discussed below.

\begin{figure}[t]
    \centering
    \includegraphics[width=\hsize,clip,page=2]{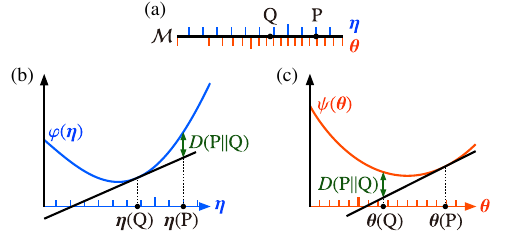}
    \caption{The definitions of the divergence $D(\P\|\Q)$. (a) Two arbitrary points $\P$, $\Q$ on the space. (b) The definition of divergence using $\bm\eta$ and $\varphi$ [Eq.~\eqref{ge_defofdiv2}]. (c) The equivalent definition using $\bm\theta$ and $\psi$ [Eq.~\eqref{ge_defofdiv3}]. }
    \label{fig_duallyflatgeometry2}
\end{figure}

\subsection{Bregman divergence and differential geometry}
\label{subsec_ge_divergence}

We introduce a two-point function called the Bregman divergence~\cite{bregman1967relaxation} based on the quadruplet $(\bm\eta,\varphi(\bm\eta),\bm\theta,\psi(\bm\theta))$, which plays a central role in dually flat geometry. Letting $\P$ and $\Q$ be two points on the space [Fig.~\ref{fig_duallyflatgeometry2}(a)], the Bregman divergence between $\P$ and $\Q$ is defined as
\begin{subequations}
\label{ge_defofdiv_all}
\begin{align}
  &D( \P \| \Q ) \notag \\
  &\!\coloneqq \varphi(\bm\eta(\P)) + \psi(\bm\theta(\Q)) -  \sum_\rho \theta^\rho(\Q) \eta_\rho(\P)
  \label{ge_defofdiv1}\\
  & = \varphi(\bm\eta(\P)) - \varphi(\bm\eta(\Q)) - \sum_\rho \theta^\rho(\Q) \qty [ \strut \eta_\rho(\P) - \eta_\rho(\Q) ] 
  \label{ge_defofdiv2} \\
  & = \psi(\bm\theta(\Q)) - \psi(\bm\theta(\P)) - \sum_\rho \eta_\rho(\P) \qty[ \strut \theta^\rho(\Q) - \theta^\rho(\P) ],
  \label{ge_defofdiv3}
\end{align}
\end{subequations}
where the equivalence of these three definitions follows from the Legendre duality~\eqref{ge_defofthetapsi} and \eqref{ge_invetavarphi}. In the second definition \eqref{ge_defofdiv2}, the terms $-\varphi(\bm\eta(\Q))-\sum_\rho \theta^\rho(\Q)\qty[ \strut \eta_\rho(\P) - \eta_\rho(\Q) ] \eqqcolon -g(\P)$ gives the tangent plane of $\varphi$ at the contact point $\Q$. Therefore, the second definition states that $D(\,\cdot\,\|\Q)=\varphi-g$ is the difference between $\varphi$ and this tangent plane $g$ [Fig.~\ref{fig_duallyflatgeometry2}(b)]. Since $\varphi$ is strictly convex, it is always larger than the tangent plane. Therefore, the divergence has the following properties
\begin{subequations}
\begin{gather}
    D(\P\|\Q) \geq 0,
    \label{ge_nonnegativityofdiv} \\
    D(\P\|\Q)=0 \iff \P=\Q.
    \label{ge_nondegeneracy}
\end{gather}
\end{subequations}
We can also interpret the Bregman divergence as a difference between $\psi$ and its tangent plane based on the third definition~\eqref{ge_defofdiv3} [Fig.~\ref{fig_duallyflatgeometry2}(c)].

\begin{figure}[t]
    \centering
    \includegraphics[width=\hsize,clip,page=3]{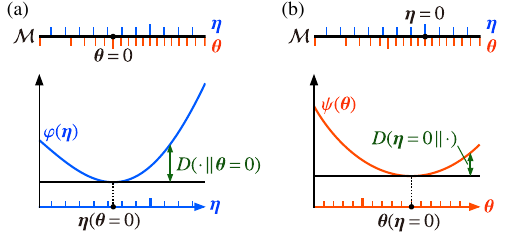}
    \caption{Two special cases of the divergence reproducing the convex functions. (a) The divergence $D(\cdot \,\| \bm\theta=0)$ reproduces the convex function $\varphi$ up to a constant. (b) The divergence $D(\bm\eta=0\|\,\cdot)$ reproduces the convex function $\psi$ up to a constant.}
    \label{fig_duallyflatgeometry3}
\end{figure}

Special cases of the divergence are when $\bm \theta(\Q) = 0$ or $\bm\eta(\P) = 0$. When the second point $\Q$ corresponds to the origin of the \utheta-coordinates, the divergence equals $\varphi$ up to the constant $\varphi(\bm\eta(\bm\theta=0))$,
\begin{equation}
    D(\bm\eta \| \bm\theta=0 ) = \varphi(\bm\eta) - \varphi(\bm\eta(\bm\theta=0)),
    \label{ge_div_varphionly}
\end{equation}
as verified from Eq.~\eqref{ge_defofdiv2} [Fig.~\ref{fig_duallyflatgeometry3}(a)]. Here we abuse the notation to use the coordinates as the arguments of the divergence, but the meaning should be clear since the points and the coordinates have the one-to-one correspondence. Similarly, when the first point $\P$ corresponds to the origin of the \ueta-coordinates, the divergence reduces to $\psi$ up to the constant $\psi(\bm\theta(\bm\eta=0))$,
\begin{equation}
    D(\bm\eta=0 \| \bm\theta) =  \psi(\bm\theta) - \psi(\bm\theta(\bm\eta=0)),
    \label{ge_div_psionly}    
\end{equation}
as verified from Eq.~\eqref{ge_defofdiv3} [Fig.~\ref{fig_duallyflatgeometry3}(b)].

Based on the Bregman divergence, we introduce a differential geometry on the space $\mathcal M$. Consider two points $\P,\Q\in\mathcal M$ that are infinitesimally close to each other. Their corresponding \ueta-coordinates are $\bm\eta$ and $\bm\eta+d\bm\eta$, and their corresponding \utheta-coordinates are $\bm\theta$ and $\bm\theta+d\bm\theta$. In this case, the Bregman divergence between $\P$ and $\Q$ is symmetric $2D(\Q\|\P) \cong 2D(\P\|\Q) \cong \sum_\rho d\theta^\rho d\eta_\rho$ up to the second order in $d\bm\eta$ and $d\bm\theta$. We write this quantity as $ds^2$,
\begin{subequations}
\label{ge_lineelement_all}
\begin{align}
    ds^2 &\!\coloneqq  \sum_\rho d\theta^\rho d\eta_\rho
    \label{ge_lineelement1} \\
    &= \sum_{\rho,\sigma} \pdv{\varphi}{\eta_\rho}{\eta_\sigma} d\eta_\rho d\eta_\sigma 
    \label{ge_lineelement2} \\
    &= \sum_{\rho,\sigma} \pdv{\psi}{\theta^\rho}{\theta^\sigma} d\theta^\rho d\theta^\sigma,
    \label{ge_lineelement3}
\end{align}
\end{subequations}
and equip the space with a Riemannian geometry by interpreting $ds^2$ as the square of the line element. 

Since the Bregman divergence between two infinitesimally close points gives the square of the infinitesimal distance, the Bregman divergence between arbitrary two points can be interpreted as a generalization of the squared distance between the two points. This generalization breaks the symmetric property, i.e., $D( \P \| \Q ) \neq D( \Q \| \P )$ in general.

\subsection{Invariance under affine transformations}
\label{subsec_ge_affine}

\begin{figure}[t]
    \centering
    \includegraphics[width=\hsize,clip,page=4]{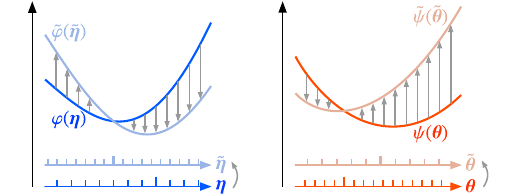}
    \caption{An affine transformation, which moves the origins of the coordinates, transforms the coordinates linearly, and adds linear functions to the convex functions [Eqs.~\eqref{ge_affine1} and \eqref{ge_affine2}]. Affine transformations keep the geometric structure invariant.}
    \label{fig_duallyflatgeometry4}
\end{figure}

We show that the geometric structure of dually flat geometry is invariant under an affine transformation~(Fig.~\ref{fig_duallyflatgeometry4}). Let $\Tns{D_\rho^\sigma}$ and $(D^{-1})\Tns{_\sigma^\rho}$ be an arbitrary $K\times K$ invertible matrix and its inverse matrix, whose product $\sum_{\rho}\Tns{(D^{-1})_\sigma^\rho D_\rho^\lambda} = \Tns{\delta_\sigma^\lambda}$ gives the Kronecker delta. Let $(a^\rho)_{\rho=1}^K$ and $(b_\rho)_{\rho=1}^K$ be arbitrary two sets of constants, and $C$ be another arbitrary constant. Using these constants, we introduce two new coordinate systems, $\tildebm\eta$ and $\tildebm\theta$, by the following affine transformation,
\begin{equation}
\begin{aligned}
&\tilde\eta_\rho(\bm\eta) = \sum_\sigma \Tns{D_\rho^\sigma}\eta_\sigma + b_\rho, \\
&\tilde\theta^\rho(\bm\theta) = \sum_\sigma \theta^\sigma \Tns{(D^{-1})_\sigma^\rho} + a^\rho.
\end{aligned}
\label{ge_affine1}
\end{equation}
We also introduce two new strictly convex functions, $\tilde\varphi(\tildebm\eta)$ and $\tilde\psi(\tildebm\theta)$, by
\begin{equation}
\begin{aligned}
&\tilde\varphi(\tildebm\eta) = \varphi(\bm\eta(\tildebm\eta))
+ \sum_\rho a^\rho\tilde\eta_\rho + C,\\
&\tilde\psi(\tildebm\theta) = \psi(\bm\theta(\tildebm\theta))
+ \sum_\rho b_\rho\tilde\theta^\rho - \sum_\rho  b_\rho a^\rho - C.
\end{aligned}
\label{ge_affine2}
\end{equation}
The new quadruplet $(\tildebm\eta,\tilde\varphi,\tildebm\theta,\tilde\psi)$ satisfies the Legendre duality in Eqs.~\eqref{ge_defofthetapsi} and \eqref{ge_invetavarphi} with the replacement of $(\bm\eta,\varphi,\bm\theta,\psi)$ by $(\tildebm\eta,\tilde\varphi,\tildebm\theta,\tilde\psi)$. Moreover, the Bregman divergence between two points $\P, \Q \in \mathcal M$ is shown to be invariant:
\begin{align}
D( \P \| \Q )  &=\varphi(\bm\eta(\P)) + \psi(\bm\theta(\Q)) - \sum_\rho \theta^\rho(\Q) \eta_\rho(\P) \notag \\
 & =  \tilde\varphi(\tildebm\eta(\P)) + \tilde\psi(\tildebm\theta(\Q)) - \sum_\rho \tilde\theta^\rho(\Q) \tilde\eta_\rho(\P).
\end{align}
Since all the quantities of the dually flat geometry are induced from the Bregman divergence, the geometric structure is invariant under an affine transformation.

\subsection{Examples}
We show two examples of dually flat geometry.

\textit{Example~1.---}
We set $\mathcal M = \mathcal V = \mathbb R^K$ and
\begin{equation}
\varphi(\bm\eta) = \frac{1}{2} \sum_\rho (\eta_\rho)^2.
\label{ge_varphiofEuclid}
\end{equation}
The other coordinates and convex function are given by
\begin{gather}
\theta^\rho(\bm\eta) = \eta_\rho , \quad \psi(\bm\theta) = \frac{1}{2} \sum_\rho \qty(\theta^\rho)^2, 
\end{gather}
with $\mathcal U = \mathbb R^K$. This geometry is self-dual, i.e., the \utheta-coordinates are the same as the \ueta-coordinates, and $\psi$ is equal to $\varphi$. The Bregman divergence is calculated as
\begin{align}
D( \P \| \Q ) 
= \frac{1}{2} \sum_\rho \qty[ \eta_\rho(\P) - \eta_\rho(\Q) ]^2,
\label{ge_Euclideandiv}
\end{align}
and the line element is $ds^2 =\sum_\rho (d\eta_\rho)^2$. Therefore, the geometry induced from this Bregman divergence is the Euclidean geometry.

\textit{Example~2.---}
We set $\mathcal M = \mathcal V = \mathbb R^K_{>0}$ and
\begin{equation}
\varphi(\bm\eta) =  \sum_\rho \qty( \strut \eta_\rho \ln \eta_\rho - \eta_\rho).
\end{equation}
The other coordinates and convex function are given by
\begin{gather}
\theta^\rho(\bm\eta) = \ln \eta_\rho , \quad \psi(\bm\theta) = \sum_\rho \exp(\theta^\rho),
\end{gather}
with $\mathcal U = \mathbb R^K$. The Bregman divergence is calculated as
\begin{align}
D( \P \| \Q ) = \sum_\rho \qty[ \eta_\rho(\P) \ln\frac{\eta_\rho(\P)}{\eta_\rho(\Q)} - \eta_\rho(\P) + \eta_\rho(\Q) ].
\label{ge_KL_divergence}
\end{align}
This Bregman divergence is called the KL divergence, which plays an essential role in information theory~\cite{cover2006elements}. We remark that some literature calls this divergence \textit{the generalized KL divergence} and saves the name \textit{KL divergence} for when $\eta_\rho$ is a probability distribution satisfying the normalization $\sum_{\rho} \eta_\rho = 1$, but we do not adopt this terminology. As discussed later, the KL divergence appears in the chemical thermodynamics of ideal dilute solutions, where $\eta_\rho$ denotes the amounts of the $\rho$th species. The line element is $ds^2 =\sum_\rho (d\eta_\rho)^2 /\eta_\rho$, which is the form of the Fisher information.

\subsection{Generalization for potentials violating the strict convexity}
\label{subsec_ge_generalization}

The dually flat geometry explained so far is restricted to those constructed from a strictly convex function $\varphi(\bm\eta)$. For later use, we extend the framework for $\varphi(\bm\eta)$ that is convex but not strictly convex (see Refs.~\cite{rockafellar1996,tasaki2022} for Legendre transformations for such functions). Consider a space $\mathcal M$ equipped with an \ueta-coordinate system and a convex function $\varphi(\bm\eta)$ that violates the strict convexity. Assume that the points $\P \in \mathcal M$ and $\bm\eta\in \mathcal V$ admit a one-to-one correspondence. We still assume the twice-differentiability of~$\varphi(\bm\eta)$.

We can still introduce $\bm\theta(\bm\eta)$ by Eq.~\eqref{ge_defoftheta} as a function of $\bm\eta$. However, the correspondence between $\bm\eta$ and $\bm\theta$ may be many-to-one. Therefore, the inverse function $\bm\eta(\bm\theta)$ cannot be defined. We can introduce $\psi(\bm\theta)$ by Eq.~\eqref{ge_defofpsi} as a function of $\bm\theta$. Although the right-hand side of \eqref{ge_defofpsi} includes the ill-defined inverse function $\bm\eta(\bm\theta)$, the right-hand side is independent of $\bm\eta$ as long as $\bm\eta$ corresponds to the same $\bm\theta$. Therefore, we can use any one of $\bm\eta$ corresponding to a $\bm\theta$ to calculate $\psi(\bm\theta)$. The convex function $\psi(\bm\theta)$ is convex but not necessarily strictly convex. Moreover, $\psi(\bm\theta)$ may lose differentiability at some $\bm\theta$. 

The inverse relation for $\bm\eta$~\eqref{ge_inveta} remains true as long as $\psi$ is differentiable. The inverse relation for $\varphi(\bm\eta)$~\eqref{ge_invvarphi} remains true without proviso. 

We can still define the Bregman divergence by Eq.~\eqref{ge_defofdiv_all}, and the equivalence of the three definitions is still guaranteed. The divergence is still non-negative \eqref{ge_nonnegativityofdiv}, but the non-degeneracy \eqref{ge_nondegeneracy} is broken. The differential geometry is constructed using the line element \eqref{ge_lineelement1}. The second expression \eqref{ge_lineelement2} is still valid, but the third one \eqref{ge_lineelement3} is valid only where $\psi(\bm\theta)$ is twice-differentiable. Moreover, since a single $\bm\theta$ may correspond to many $\bm\eta$, a nonzero displacement $d\bm\eta$ may be associated with $d\bm\theta=0$. For such displacements, we have $ds^2=0$.

The invariance under an affine transformation is completely valid. All the statements in Sec.~\ref{subsec_ge_affine} hold as they are.

We do not discuss these properties more concretely in this section. Instead, we will use this generalization in Sec.~\ref{sec_open} to construct a dually flat geometry for open chemical reaction systems and discuss the consequences of the lack of strict convexity on the constructed geometry.

\section{Dually flat geometry for closed chemical reaction systems}
\label{sec_closed}

We construct a dually flat geometry for closed chemical reaction systems. Our strategy is to take linear combinations of the distribution $\bm n$ as the \ueta-coordinates and the Gibbs free energy $G(\bm n)$ as the convex function $\varphi$. In constructing the \ueta-coordinates, we take into account the constraint due to the constitutive equation~\eqref{ch_constitutive} to reflect the reactions in the system.

\subsection{Space}

\begin{figure}[b]
    \centering
    \includegraphics[width=\hsize,clip,page=1]{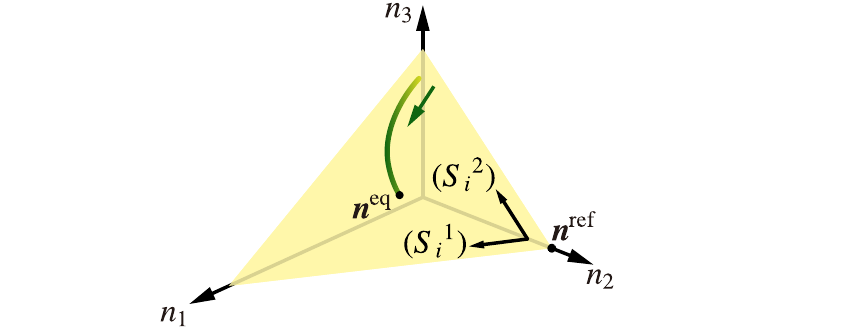}
    \caption{An example of a positive stoichiometric compatibility class [Eq.~\eqref{cl_scclass}], the space reachable from a reference state $\bm n^\refr$ with a linear combination of the reactions $(S{}_i{}^1),\dots, (S{}_i{}^M)$. A positive stoichiometric compatibility class usually contains a unique equilibrium state $\bm n^\eq$ in it, and typical trajectories (exemplified by the green line) relax to the equilibrium.}
    \label{fig_geometry_space}
\end{figure}

We use a \textit{positive stoichiometric compatibility class}~\cite{feinberg2019foundations} as the space (Fig.~\ref{fig_geometry_space}). A positive stoichiometric compatibility class is the set of positive distributions reachable from an arbitrary reference state $\bm{n}^\refr \equiv (n_i^\refr)_{i=1}^N$ through the reactions:
\begin{equation}
\mathcal M(\bm n^\refr) \coloneqq \Set{ \qty(n^\refr_i + \sum_\rho\Tns{S_i^\rho} \xi_\rho)_{i=1}^N | (\xi_\rho)_{\rho=1}^M\in\mathbb{R}^M } \cap \mathbb{R}_{> 0}^N.
 \label{cl_scclass}
\end{equation}
By taking the intersection with $\mathbb R^N_{>0}$, we impose $n_i>0$ for all the species since the chemical potential may be ill-defined for $n_i=0$. Two reference states $\bm n^{\refr\,(1)}, \bm n^{\refr\,(2)}$ induce the same space $\mathcal M(\bm n^{\refr\,(1)}) = \mathcal M(\bm n^{\refr\,(2)})$ if they satisfy $n^{\refr\,(1)}_i + \sum_\rho\Tns{S_i^\rho} \xi_\rho = n^{\refr\,(2)}_i$ for some $(\xi_\rho)_{\rho=1}^M$. We construct a dually flat geometry on $\mathcal M(\bm n^\refr)$ with an arbitrary $\bm n^\refr$, and we fix $\bm n^\refr$ throughout the construction. 

A time evolution of a closed system is likely to be confined in a single positive stoichiometric compatibility class since the variation of the amounts is a linear combination of $(\Tns{S_i^\rho})_{i=1}^N$ due to the constitutive equation \eqref{ch_constitutive}. The exception is when some of the amounts become zero during a process, but such a situation arises only for pathological kinetics.

Let $K$ denote the dimensionality of the positive stoichiometric compatibility class. It equals the number of independent reactions, i.e., the number of linearly independent columns of the stoichiometric matrix. Therefore, we always have $K\leq M$, where $M$ is the number of the reactions.

\subsection{Reduced stoichiometric matrix}

In preparation for constructing \ueta-coordinates, we introduce a \textit{reduced stoichiometric matrix} $\mathsfit A$ by discarding the linearly-dependent reactions from the stoichiometric matrix $\mathsfit S$. Concretely, we take an arbitrary linearly independent basis of the $K$-dimensional linear space
\begin{equation}
 \Im \mathsfit S =  \Set{ \qty( \sum_\rho\Tns{S_i^\rho} \xi_\rho)_{i=1}^N | (\xi_\rho)_{\rho=1}^M\in\mathbb{R}^M },
\end{equation}
where Im denotes the image of a matrix. We write the basis vectors as $(\Tns{A_i^1}){}_{i=1}^N,\,\allowbreak  (\Tns{A_i^2}){}_{i=1}^N, \,\allowbreak  \dots,\,\allowbreak  (\Tns{A_i^K}){}_{i=1}^N$ and call the $N\times K$ matrix $\mathsfit A \equiv (\Tns{A_i^\rho}){}_{i=1}^N{}_{\rho=1}^K$ a \textit{reduced stoichiometric matrix}. Since we have the freedom of changing the basis, $\mathsfit A$ is not unique. For example, $\mathsfit A$ can be constructed simply by eliminating $M-K$ columns of $\mathsfit S$ that are linear dependent on the remaining $K$ columns. The choice of the matrix $\mathsfit A$ does not affect our construction of dually flat geometry below. 

For an intuitive understanding of $\mathsfit A$, we introduce fictitious reactions
\begin{equation}
    \Tns{{\tilde\lambda}_1^\rho} \X_1 + \dots + \Tns{{\tilde\lambda}_N^\rho} \X_N
    \,\rightleftarrows\,
    \Tns{{\tilde\kappa}_1^\rho} \X_1 + \dots + \Tns{{\tilde\kappa}_N^\rho} \X_N,
    \label{cl_reduced_reaction}
\end{equation}
for $\rho=1,\dots,K$, where $\Tns{{\tilde\lambda}_i^\rho}$ and $\Tns{{\tilde\kappa}_i^\rho}$ are defined by $\Tns{{\tilde\lambda}_i^\rho} \coloneqq \max \{ -{\Tns{A_i^\rho}} , 0\}$ and $\Tns{{\tilde\kappa}_i^\rho}\coloneqq \max \{ {\Tns{A_i^\rho}} , 0\} $ so that $\Tns{A_i^\rho} = \Tns{{\tilde\kappa}_i^\rho}- \Tns{{\tilde\lambda}_i^\rho} $. We call this reaction the $\rho$\textit{th reduced reaction}. The use of $\mathsfit A$ amounts to considering a virtual reaction system with these reduced reactions.

Each reduced reaction can be realized as a combination of the actual reactions, and conversely, each actual reaction is uniquely interpreted as a combination of the reduced reactions. Indeed, a column of $\mathsfit S$ is uniquely expressed as a linear combination of the columns of $\mathsfit A$ as
\begin{equation}
  \Tns{S_i^\sigma} = \sum_{\rho=1}^{K}  \Tns{A_i^\rho}\Tns{c_\rho^\sigma},
  \label{cl_S_Sc}
\end{equation}
since the columns of $\mathsfit A$ form a basis of $\Im \mathsfit S$. Here, the $K\times M$ matrix $(\Tns{c_\rho^\sigma})$ is of full rank (rank $K$).

\vspace*{\baselineskip}
\textit{Example system 1}.---%
For a demonstration throughout this section, we use the following model system with five species $\X_1,\dots,\X_5$, among which $\X_1$ is the solvent species and the others are the solute species:
\begin{equation}
\begin{alignedat}{2}
&\rho=1: &\quad& 2\X_2 \,\rightleftarrows\, \X_3,\\
&\rho=2: &\quad& 2\X_2 + \X_5  \,\rightleftarrows\, \X_3 + \X_5 ,\\
&\rho=3: &\quad& \X_2 + \X_3 \,\rightleftarrows\, \X_4.
\end{alignedat}
\label{cl_example1}
\end{equation}
This system, in effect, converts three molecules of $\X_2$ into one molecule of $\X_4$. The species $\X_3$ is a reaction intermediate, and $\X_5$ is a catalyst. We do not necessarily assume the ideal dilute property for this example system.

The positive stoichiometric compatibility class for the example system is two-dimensional since the first and the second reactions are redundant. Therefore, we can construct a reduced stoichiometric matrix by simply removing the second column from the original stoichiometric matrix. However, we can further simplify the reduced stoichiometric matrix by making some linear combinations of the reactions. For example, we can choose
\begin{equation}
\begin{alignedat}{2}
&\rho=1: &\quad& 2\X_2 \,\rightleftarrows\, \X_3,\\
&\rho=2: &\quad& 3\X_2  \,\rightleftarrows\, \X_4.
\end{alignedat}
\end{equation}
as the reduced reactions. The matrices $\mathsfit S = (\Tns{S_i^\rho})$, $\mathsfit A = (\Tns{A_i^\rho})$, $\mathsfit c = (\Tns{c_\rho^\sigma})$ are given by
\begin{equation}
    \mathsfit S = \mqty(0 & 0 & 0 \\ -2 & -2 & -1 \\ 1 & 1 & -1 \\ 0 & 0 & 1 \\ 0 & 0 & 0 ),\quad 
    \mathsfit A = \mqty(0 & 0\\ -2 & -3 \\ 1 & 0 \\0 & 1\\ 0& 0), \quad 
    \mathsfit c = \mqty(1 & 1 & -1 \\ 0 & 0 & 1).
\end{equation}
We use this choice of reduced reactions for the demonstration throughout.

\subsection{Coordinates and the convex functions}
\label{subsec_cl_coordinates}

\subsubsection{Construction}

We construct an \ueta-coordinate system as linear combinations of the amounts of substances. Since the columns of $\mathsfit A$ form a basis of $\Im \mathsfit S$, any point $\bm n \in \mathcal M(\bm n^\refr)$ admits a unique expression 
\begin{equation}
n_i = n_i^\refr + \sum_\rho \Tns{A_i^\rho}\eta_\rho.
\label{cl_defofeta}
\end{equation}
We use this equation for the definition of the \ueta-coordinates. The range $\mathcal V$ of $\bm\eta$ is taken so that the corresponding $n_i$ are all positive. We have the one-to-one correspondence between $\bm\eta\in \mathcal V$ and the distributions $\bm n \in \mathcal M(\bm n^\refr) $.

We take the Gibbs free energy as the convex function $\varphi(\bm\eta) \coloneqq G(\bm n(\bm \eta))$.  The Gibbs free energy is strictly convex within $\mathcal M(\bm n^\refr)$, verified as follows. Recall that we have assumed that the inequality~\eqref{ch_convexity} of the convexity of $G(\bm n)$ is saturated only when $\bm n\propto \bm n'$. Since we have also assumed that the reactions conserve the total mass, two distributions $\bm n \neq \bm n'$ with $\bm n\propto \bm n'$ cannot reside in a single $\mathcal M(\bm n^\refr)$. Therefore, the inequality is never saturated for $\bm n, \bm n'\in \mathcal M(\bm n^\refr)$.

We construct $\bm\theta$ and $\psi$ from the definition~\eqref{ge_defofthetapsi}. As a result, we obtain the following quadruplet of dually flat geometry:
\begin{subequations}
\label{cl_quadruplet_all}
\begin{empheq}[left={\empheqlbrace}\hspace{0.4em}]{align=1}
& \eta_\rho(\bm n) = \sum_i (A^-)\Tns{_\rho^i}(\strut n_i-n_i^\refr) ,
\label{cl_etaofn}\\
& \varphi(\bm\eta(\bm n)) = G \qty(\bm n) = \sum_i \mu^i(\bm n) n_i,
\label{cl_varphi}\\
&   \theta^\rho(\bm n) = \sum_i \mu^i(\bm n) \Tns{A_i^\rho} ,
\label{cl_theta} \\
& \psi(\bm\theta(\bm n)) = 
- \sum_i \mu^i(\bm n) n_i^\refr.
\label{cl_psi}
\end{empheq}
\end{subequations}
This quadruplet completely characterizes our information-geometric structure. Here, Eq.~\eqref{cl_etaofn} is obtained by inverting the definition \eqref{cl_defofeta}, and $(A^-)\Tns{_\rho^i}$ is a pseudo-inverse matrix of $\Tns{A_i^\rho}$, defined by an arbitrary matrix satisfying $\mathsfit A \mathsfit A^- \mathsfit A = \mathsfit A$.
Pseudo-inverse matrices of $\mathsfit A$ are not unique, but the right-hand side of Eq.~\eqref{cl_etaofn} does not depend on this ambiguity. See appendix \ref{sec_appendix_pinv} for properties of the pseudo-inverse matrices.

We have left some arbitrariness in the above-constructed quadruplet, but the arbitrariness is understood as the freedom of the affine transformation [Eqs.~\eqref{ge_affine1} and \eqref{ge_affine2}] and therefore do not affect the induced geometric structures. Indeed, the arbitrariness in the reduced stoichiometric matrix $\mathsfit A$, i.e., the choice of the reduced reactions, corresponds to the freedom of affine transformation with a regular matrix $(\Tns{D_\rho^\sigma})$ in \eqref{ge_affine1}. The arbitrariness of the reference point $\bm n^\refr$ corresponds to the affine transformation with $(b_\rho)$ in Eqs.~\eqref{ge_affine1} and \eqref{ge_affine2}.

\subsubsection{Properties and interpretations}

\begin{figure}
    \centering
    \includegraphics[width=\hsize,clip,page=2]{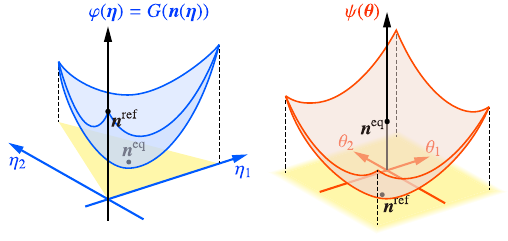}
    \caption{Schematics of the dually flat geometry for closed systems. The positive stoichiometric compatibility class (Fig.~\ref{fig_geometry_space}) is equipped with two coordinate systems, $\bm\eta$ and $\bm\theta$ (not shown). The \ueta-coordinates $\bm\eta$ is accompanied by the convex function $\varphi(\bm\eta)$, i.e., the Gibbs free energy. Their Legendre transform gives the \utheta-coordinates $\bm\theta$ and the convex function $\psi(\bm\theta)$.}
    \label{fig_geometry_close}
\end{figure}

The followings are some properties and interpretations of the quadruplet (see Fig.~\ref{fig_geometry_close}). From Eq.~\eqref{cl_defofeta}, the \ueta-coordinate $\eta_\rho$ is interpreted as the extent of the $\rho$th reduced reaction, measured from the reference point $\bm n^\refr$. In particular, the origin corresponds to the reference point:
\begin{equation}
     \eta_\rho(\bm n^\refr) = 0.
\end{equation}
This implies,  from the inverse Legendre transformation \eqref{ge_inveta}, that the function $\psi(\bm\theta)$ takes its minimum at $\bm n^\refr$.

On the other hand, the \utheta-coordinate is interpreted as the negative of the affinity of the $\rho$th reduced reaction, understood by comparing the \utheta-coordinates \eqref{cl_theta} with the definition of affinity~\eqref{ch_affinity}. Since $\bm\theta$ is in one-to-one correspondence with $\bm n \in \mathcal M(\bm n^\refr)$, we can use this set of affinities to specify a state on $\mathcal M(\bm n^\refr)$. The affinity of the original reactions $F^\rho$ and of the reduced reactions $-\theta^\rho$ are related by 
\begin{equation}
    F^\sigma = - \sum_\sigma \theta^\rho \Tns{c_\rho^\sigma},
    \label{cl_F_theta}
\end{equation}
where we have used Eq.~\eqref{cl_S_Sc}. Thus, the origin $\theta^\rho=0$ corresponds to the distribution with vanishing affinity, i.e., the equilibrium distribution $\bm n^\eq$:
\begin{equation}
    \theta^\rho(\bm n^\eq)=0.
    \label{cl_theta_origin}
\end{equation}
Since the matrix $(\Tns{c_\rho^\sigma})$ is full-rank, $F^\rho=0$ for all $\rho$ implies $\theta^\rho = 0 $ for all $\rho$ in Eq.~\eqref{cl_F_theta}. Therefore, the equilibrium state is unique within the positive stoichiometric compatibility class. Due to the Legendre transformation \eqref{ge_defoftheta}, the convex function $\varphi$ takes its minimum at the equilibrium. In other words, the Gibbs free energy takes its minimum at $\bm n^\eq$ within the positive stoichiometric compatibility class.

\begin{figure}
    \centering
    \includegraphics[width=\hsize,clip,page=3]{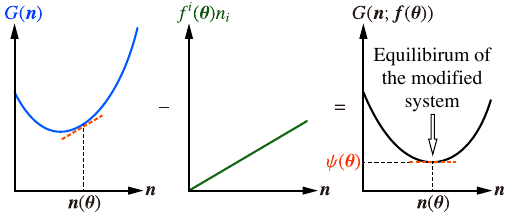}
    \caption{Schematics of the interpretation of $\bm n(\bm\theta)$ and $\psi(\bm\theta)$ in terms of `external field' $\bm f(\bm\theta)$. When the distribution $\bm n(\bm\theta)$ is a nonequilibrium distribution, the nonzero slope of $G(\bm n)$ gives the nonzero affinity. If we apply an external field $-\sum_i f^i(\bm\theta) n_i$ to the system to cancel out the slope, the distribution $\bm n(\bm\theta)$ becomes an equilibrium of the modified system. The equilibrium free energy of this modified system equals the potential $\psi(\bm\theta)$.}
    \label{fig_externalfield}
\end{figure}

The convex function $\psi$ is a newly introduced thermodynamic potential, and its interpretation requires some complicated discussion. For this purpose, we start with a slightly different interpretation of the \utheta-coordinates. Let us fix $\bm\theta$ for now. Its corresponding distribution $\bm n(\bm\theta)$ is a nonequilibrium distribution in general. However, if we consider applying a fictitious `external field' to the system to cancel out the affinity $-\theta^\rho$, the distribution $\bm n(\bm\theta)$ turns into an equilibrium (Fig.~\ref{fig_externalfield}). Concretely, we consider an external field $\bm f(\bm\theta) \equiv \qty(f^i(\bm\theta))_{i=1}^N$ that modifies the energy of the species $\X_i$ by $-f^i(\bm\theta)$. The modified Gibbs free energy under the external field is
\begin{equation}
    G(\bm n; \bm f) \coloneqq G(\bm n) - \sum_i f^i n_i =  \sum_i \qty[ \mu^i(\bm n) - f^i] n_i.
    \label{cl_modGibbs}
\end{equation}
To cancel out the affinity $-\theta^\rho$, the external field must satisfy
\begin{equation}
    \sum_i f^i(\bm\theta) \Tns{A_i^\rho} = \theta^\rho.
    \label{cl_thetaofv}
\end{equation}
The general solution of Eq.~\eqref{cl_thetaofv} is
\begin{equation}
    f^i(\bm\theta) = \sum_\rho \theta^\rho (A^-)\Tns{_\rho^i} + l^i(\bm\theta),
    \label{cl_voftheta}
\end{equation}
where $\bm l\equiv (l^i)_{i=1}^N$ is an arbitrary \textit{conservation law}, i.e., an arbitrary vector satisfying $\sum_i l^i \Tns{A_i^\rho} = 0$. Now let us (partially) fix this arbitrariness by further imposing
\begin{equation}
    \sum_i f^i(\bm\theta) n_i^\refr = 0.
    \label{cl_v_cond}
\end{equation}
This condition can always be fulfilled by multiplying $l^i$ by a proper $i$-independent constant. This condition may not completely fix the arbitrariness in Eq.~\eqref{cl_voftheta}. Even if so, we choose an arbitrary $\bm f$ satisfying Eqs.~\eqref{cl_voftheta} and \eqref{cl_v_cond} and write it as $\bm f(\bm\theta)$. Note that $\bm l$ may depend on $\bm\theta$ in general.

This discussion gives a variational formula to calculate $\bm n(\bm\theta)$ from a given $\bm\theta$. From the construction of the external field, $\bm n(\bm\theta)$ is the equilibrium under the modified Gibbs free energy~\eqref{cl_modGibbs}. Since the Gibbs free energy is minimum at the equilibrium, we obtain the following expression
\begin{equation}
    \bm n(\bm\theta) = \argmin_{\bm n \in \mathcal M(\bm n^\refr)} G(\bm n; \bm f(\bm \theta)).
    \label{cl_noftheta} 
\end{equation}
In other words, the \utheta-coordinates specify a distribution $\bm n$ through the external field~\eqref{cl_voftheta}--\eqref{cl_v_cond} required to reproduce $\bm n$ as an equilibrium.

We can readily explain the interpretation of $\psi(\bm\theta)$. The convex function $\psi(\bm\theta)$ is the negative of the Gibbs free energy at the equilibrium of the modified system $\bm n(\bm\theta)$:
\begin{equation}
    \psi(\bm\theta) = - \min_{\bm n \in \mathcal M(\bm n^\refr)} G(\bm n; \bm f(\bm \theta)) = -G(\bm n(\bm\theta); \bm f(\bm\theta)).
    \label{cl_psioftheta} 
\end{equation}
The proof of this formula is a direct calculation:
\begin{align}
    &\psi(\bm\theta) = \bm\theta \cdot \bm\eta(\bm\theta) - G(\bm n(\bm\theta)) \notag \\
    &= \sum_{i,\rho,j} f^i(\bm\theta) \Tns{A_i^\rho} (A^-)\Tns{_\rho^j} (n_j(\bm\theta) - n^\refr_j)  - \sum_i \mu^i(\bm n(\bm\theta)) n_i(\bm\theta) \notag \\
    &= \sum_i f^i(\bm\theta) (n_i(\bm\theta) - n^\refr_i)  - \sum_i \mu^i(\bm n(\bm\theta)) n_i(\bm\theta) \notag \\
    &= - G(\bm n(\bm\theta); \bm f(\bm\theta)),
    \label{cl_psiofGv_proof}
\end{align}
where we have used the Legendre transformation~\eqref{ge_defofpsi}, the definition of $\bm\eta$~\eqref{cl_etaofn}, Eqs.~\eqref{cl_modGibbs}--\eqref{cl_v_cond}, and a property of the pseudo-inverse matrix \eqref{ps_pinv2}. In appendix \ref{sec_appendix_variational}, we discuss another derivation of Eqs.~\eqref{cl_noftheta} and \eqref{cl_psioftheta} from the variational formula of Legendre transformation. 

We remark that the minimum of Gibbs free energy, i.e., the equilibrium distribution is not always analytically solvable, even if the system is an ideal dilute solution. Therefore, the function $\bm n(\bm\theta)$ and $\psi(\bm\theta)$ is not always analytically given from Eqs.~\eqref{cl_noftheta} and \eqref{cl_psioftheta} as functions of $\bm\theta$. Still, these equations provide a physical interpretation of these functions and a way to numerically calculate their explicit values.

\vspace*{\baselineskip}
\textit{Example system 1}.---%
For our example system introduced in \eqref{cl_example1}, the reduced stoichiometric matrix and one of its pseudo inverse are given by
\begin{equation}
    \mathsfit A = \mqty(0 & 0\\ -2 & -3 \\ 1 & 0 \\0 & 1\\ 0& 0), \quad
    \mathsfit A^- = \mqty(0 & 0 & 1 & 0 & 0 \\ 0 & 0 & 0 & 1 & 0 ).
\end{equation}
For demonstration, we choose a reference point where all the reactive solutes $\X_2,\X_3,\X_4$ are converted to $\X_2$, i.e., $\bm n^\refr=(n_1^\refr, n_2^\refr, 0, 0, n_5^\refr)$. Note that, although we restrict the geometry to the \textit{positive} stoichiometric compatibility class, we can choose a reference point with zeros. Therefore, the quadruplet is 
\begin{equation}
\left\{ \hspace{0.2em}
\begin{aligned}
& \eta_1(\bm n) = n_3 , \quad 
\eta_2(\bm n) = n_4, \\
& \varphi(\bm\eta(\bm n)) = G \qty(\bm n) = \sum_{i=2,3,4} \mu^i(\bm n) n_i +  \sum_{i=1,5} \mu^i(\bm n) n_i^\refr,\\
& \theta^1(\bm n) = \mu^3(\bm n) - 2\mu^2(\bm n), \quad
\theta^2(\bm n) = \mu^4(\bm n) - 3\mu^2(\bm n), \\
& \psi(\bm\theta(\bm n)) = -\mu^1(\bm n) n_1^\refr -\mu^2(\bm n) n_2^\refr -\mu^5(\bm n) n_5^\refr,
\end{aligned}
\right.
\end{equation}
where we have used $n_1 = n_1^\refr$ and $ n_5 = n_5^\refr$ since the amounts $n_1$ and $n_5$ do not change under the reactions.

The \ueta-coordinates are the extent of the reduced reactions measured from the reference point $\bm n^\refr$. Due to the choice of $\bm n^\refr$, the amounts $n_3$ and $n_4$ directly represent the extent of the reduced reactions. The \utheta-coordinates are the negative of the affinities.

To calculate the inverted expression $\bm n(\bm\theta)$ and $\psi(\bm\theta)$, we introduce the external field $\bm f(\bm\theta)$ as in Eq.~\eqref{cl_voftheta}:
\begin{equation}
     \mqty(f^1(\bm\theta)\\ f^2(\bm\theta)\\ f^3(\bm\theta)\\ f^4(\bm\theta)\\ f^5(\bm\theta)) = \mqty(0\\0\\\theta^1\\\theta^2\\0) + \bm l(\bm\theta),
\end{equation}
where $\bm l$ is a conservation law satisfying the constraint \eqref{cl_v_cond}
\begin{equation}
    f^1n_1^\refr + f^2 n_2^\refr + f^5 n_5^\refr = 0.
\end{equation}
In this example, the choice $\bm l = 0$ will satisfy the constraint. Therefore, the functional dependence $\bm n(\bm\theta)$ and $\psi(\bm\theta)$ is given by the minimum of the modified Gibbs free energy
\begin{equation}
    G(\bm n; \bm f(\bm\theta)) = G(\bm n) - \theta^1 n_3 - \theta^2 n_4.
\end{equation}
In this modification, $\theta^1$ and $\theta^2$ modify the energy of $\X_3$ and $\X_4$, respectively, to achieve the equilibrium of $2\X_2\rightleftarrows\X_3$ and $3\X_2\rightleftarrows\X_4$ under the nonequilibrium distribution $\bm n$.

If we assume the ideal-dilute property, we can calculate the minimum of $G(\bm n;\bm f(\bm\theta))$ explicitly. We omit the result here since the calculation requires solving a cubic equation, and the result is complicated.

\subsection{Bregman divergence and differential geometry}

The Bregman divergence between arbitrary two distributions $\bm n$, $\bm n'$ is calculated from the definition~\eqref{ge_defofdiv2} as
\begin{align}
D(\bm n\| \bm n') &= G(\bm n ) - G(\bm n') - \sum_i \mu^i(\bm n') (n_i - n_i') \notag \\
& =\sum_i \qty[ \mu^i(\bm n) - \mu^i(\bm n') ]\, n_i,
\label{cl_divergence}
\end{align}
where we have used the property of $\mathsfit A$~\eqref{ps_pinv2} in the first equality and the Euler relation~\eqref{ch_Euler} in the second equality. This expression of divergence is one of the key results in this paper. In particular, the divergence between an arbitrary state $\bm n$ and the equilibrium state $\bm n^\eq$ is
\begin{equation}
    D(\bm n\| \bm n^\eq) = G(\bm n) - G(\bm n^\eq),
    \label{cl_divergence_eq}
\end{equation}
where we have used $\bm\theta(\bm n^\eq)=0$ and Eq.~\eqref{ge_div_varphionly}.

The information-geometric line element, i.e., the squared distance between two infinitesimally close points, is calculated from the definition \eqref{ge_lineelement_all} as 
\begin{equation}
\begin{aligned}
    ds^2 &= \sum_\rho d\theta^\rho d\eta_\rho = \sum_i d\mu^i dn_i \\
    &=  \sum_{i,j} \chi^{ij} dn_i dn_j 
    = \sum_{\rho,\sigma} \qty(\sum_{i,j} \chi^{ij} \Tns{A_i^\rho}\Tns{A_j^\sigma}) d\eta_\rho d\eta_\sigma 
\end{aligned}
\label{cl_lineelement}
\end{equation}
where 
\begin{equation}
    \chi^{ij}\coloneqq \frac{\partial \mu^i}{\partial n_j} = \frac{\partial^2 G}{\partial n_i\partial n_j}
\end{equation}
is the susceptibility and equal to the Hessian of the Gibbs free energy. Therefore, the differential geometry induced from our quadruplet uses the Hessian of the Gibbs free energy as the metric.

\vspace*{\baselineskip}
\textit{Ideal dilute solutions}.---%
We explicitly calculate the divergence and the metric for ideal dilute solutions using the expression of the chemical potentials~\eqref{ch_mu_ideal_all}. The divergence is calculated as
\begin{equation}
    D(\bm n \| \bm n')
    = n_1RT \sum_{i=2}^N \qty( x_i \ln \frac{x_i}{x_i'} - x_i + x'_i),
    \label{cl_divergence_ideal}    
\end{equation}
where we have introduced the mole fraction of the solutes $x_i \coloneqq n_i/ n_1$ and $x'_i \coloneqq n'_i/ n'_1$ for $i=2,\dots, N$. This expression holds even if $n_1 \neq n'_1$, i.e., when the solvent reacts with the solutes.

This expression coincides with the KL divergence~\eqref{ge_KL_divergence} between the two distributions of the mole fractions of the solutes, up to the factor $n_1RT$. In particular, the difference of Gibbs free energy $G(\bm n) - G(\bm n^\eq)$ [Eq.~\eqref{cl_divergence_eq}] is given by the KL divergence between $x_i$ and $x_i^\eq \coloneqq n_i^\eq/n_1^\eq$. The latter coincidence has already been pointed out, for example, in Refs.~\cite{shear1967analog, higgins1968some,horn1972general,rao2016nonequilibrium, ge2016nonequilibrium, ge2016mesoscopic, yoshimura2021information}. From this viewpoint, our divergence for possibly non-ideal solutions~\eqref{cl_divergence} is regarded as a natural generalization of the KL divergence.

The line element is
\begin{equation}
    ds^2 = n_1 RT \sum_{i=2}^N \frac{1}{x_i} (dx_i)^2,
\end{equation}
which holds even if $dn_1\neq 0$. This form is similar to the Fisher information in  stochastic thermodynamics. This similarity has been used to derive a geometrical speed limit in previous work \cite{yoshimura2021information}.

\subsection{Relation to the flows}

Although we mainly focus on distributions and the constitutive equations to construct the dually flat geometry (Fig.~\ref{fig_flow_distribution}), the geometry possesses some natural relations to flows and kinetics. First, we calculate how flows determine the variation of the \ueta-coordinates. Using the expression of $n_i$ by the \ueta-coordinates~\eqref{cl_defofeta} and the expression of $dn_i/dt$ by the flows \eqref{ch_constitutive}, we find
\begin{equation}
    \sum_\rho \Tns{A_i^\rho} \frac{d\eta_\rho}{dt} 
    = \frac{dn_i}{dt}
    = \sum_\rho \Tns{S_i^\rho} J_\rho(t)
    = \sum_{\rho,\sigma} \Tns{A_i^\rho}\Tns{c_\rho^\sigma} J_\sigma(t),
\end{equation}
where we have inserted Eq.~\eqref{cl_S_Sc} in the last equality.
Since all the columns of $\mathsfit A$ are linearly independent, we can eliminate $\mathsfit A$ from the first and the last side to obtain
\begin{equation}
  \frac{d\eta_\rho}{dt} = \sum_{\sigma=1}^{M} \Tns{c_\rho^\sigma} J_\sigma(t).
  \label{cl_etatoJ}
\end{equation}
This equation gives the change of \ueta-coordinates in terms of the flows of the actual reactions.

Second, and more importantly, the entropy production rate $\dot\Sigma$ is expressed by geometric quantities introduced above, and therefore the time evolutions of the quantities are constrained by the second law of thermodynamics $\dot\Sigma \geq 0$. Recall that the entropy production rate is proportional to the time derivative of the Gibbs free energy in closed systems [Eq.~\eqref{ch_sigma_G}]. Since the convex function $\varphi$ is the Gibbs free energy, we have
\begin{equation}
    T\dot\Sigma = - \frac{d\varphi}{dt} \geq 0.
    \label{cl_epr1}
\end{equation}
From the definition of the \utheta-coordinates as a derivative of $\varphi$ [Eq.~\eqref{ge_defoftheta}], we can rewrite Eq.~\eqref{cl_epr1} as
\begin{equation}
    T \dot\Sigma = -\sum_{\rho} \theta^\rho \frac{d\eta_\rho}{dt} \geq 0.
    \label{cl_epr2}
\end{equation}
In other words, since $d\eta_\rho/dt$ is the time derivative of the extent of the $\rho$th reduced reaction, and $-\theta^\rho$ is the negative of its affinity, their product gives the entropy production rate. Moreover, using the relation between the divergence and the Gibbs free energy~\eqref{cl_divergence_eq}, we obtain
\begin{equation}
    T\dot\Sigma = -\frac{d}{dt}D(\bm n\| \bm n^\eq) \geq 0.
    \label{cl_epr3}
\end{equation}
Thus, the second law of thermodynamics implies that $D(\bm n\| \bm n^\eq)$ is a Lyapunov function.

\section{Dually flat geometry for open chemical reaction systems}
\label{sec_open}

\subsection{Outline}
\label{subsec_op_outline}

This section constructs a dually flat geometry for open chemical reaction systems. In open systems, the Gibbs free energy no longer equals the cumulative entropy production. Instead, we use the cumulative entropy production itself as the convex function $\varphi$. Since entropy production is not solely determined by the change in the amounts in the system, we need to use a broader space with which we can track all the entropy production.

To naturally find such a space, we utilize the specific realization of open reaction systems using particle reservoirs (Sec.~\ref{subsec_ch_reservoir}, Fig.~\ref{fig_systems_reservoir}). We can regard the system and the reservoirs together as a closed reaction system, and its total Gibbs free energy gives the cumulative entropy production. We apply the dually flat geometry for closed systems to the total system to construct a dually flat geometry for open systems. 

In the construction, we divide the reduced reactions into two types of reactions: the `change' reactions, which change the amounts of closed species, and the `emergent cycle'~\cite{polettini2014irreversible} reactions, which only involve open species. Since the reduced reactions label the coordinates, this division leads to the division of the coordinates as $\bm \eta\equiv(\bm\eta_\chg, \bm\eta_\cyc)$ and $\bm \theta\equiv(\bm\theta^\chg, \bm\theta^\cyc)$, where `$\chg$' stands for `change,' and `$\cyc$' stands for `cycle.'

Although the construction exploits the specific realization of an open system with particle reservoirs, the resulting geometry is applicable for any realization of a potentiostatted open system. Indeed, we can alternatively introduce the same geometry without relying on the particle reservoirs but using the cumulative entropy production calculated from the cumulative flows (Sec.~\ref{subsec_op_flow}).

The construction in Secs.~\ref{subsec_op_space}--\ref{subsec_op_coordinates} is rather technical. One can directly jump to Fig.~\ref{fig_geometry_open} and its caption for the idea of the geometry for open reaction systems and to Sec.~\ref{subsec_op_divergence} for the resulting divergence and the line element.

\subsection{Space}
\label{subsec_op_space}

\begin{figure}
    \centering
    \includegraphics[width=\hsize,clip,page=4]{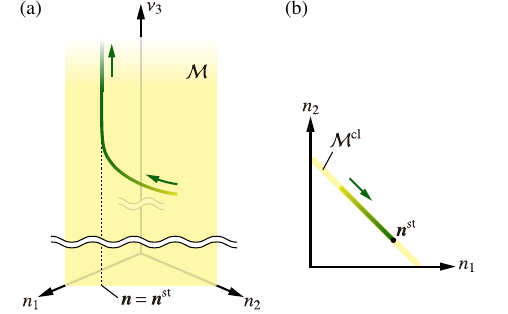}
    \caption{Schematics of the spaces for an open system. A trajectory relaxing toward a steady state $\bm n^\st$ is drawn in each space (green curve). (a) The space of the total amounts in the system and the particle reservoirs together $(\bm n, \bm\nu)$. The wavy line indicates that the amounts of the open species $\bm\nu$ are vast. The amounts $\bm\nu$ may continue to evolve even when the principal system is in the steady state. In the schematics, we omit at least one open species (say $\nu_4$) that decreases along the trajectory to compensate for the increase in $\nu_3$ to conserve the total mass.
    (b) The corresponding space of the amounts of the closed species in the principal system $\bm n$. The trajectory terminates at the steady state.}
    \label{fig_spaces_open}
\end{figure}

The positive stoichiometric compatibility class~\eqref{cl_scclass} for the total system is [Fig.~\ref{fig_spaces_open}(a)]
\begin{align}
&\mathcal M (\bm n^\refr, \bm\nu^\refr)
= \Set{ \mqty( \bm n^\refr \\ \bm\nu^\refr)  + \mathsfit S \bm \xi  |  \bm \xi \in \mathbb R^M}  \cap \mathbb R^N_{> 0} .
\label{op_scclass_tot}
\end{align}
Even if the principal system is in a steady state $\bm n = \bm n^\st$, the total amounts $\bm\nu$ may continue to evolve, and therefore the trajectory may extend infinitely.

By discarding $\bm\nu$, we can also consider a narrower space [Fig.~\ref{fig_spaces_open}(b)]: 
\begin{equation}
\mathcal M^\cl(\bm n^\refr) 
\coloneqq \Set{  \bm n^\refr + \mathsfit S_\cl \bm \xi | \bm\xi \in \mathbb R^{M}}
\cap \mathbb R^{N_\cl}_{> 0} ,
\label{op_scclass_cl} 
\end{equation}
where $\mathsfit S_\cl = (\Tns{S_i^\rho})_{i=1}^{N_\cl} \,{}_{\rho=1}^M$ is the first $N_\cl$ rows of $\mathsfit S$. We do not construct a dually flat geometry on $\mathcal M^\cl$, but we will use it for some discussions. In this space, trajectories terminate at a steady state.

The space $\mathcal M$ has the same or higher dimensionality than $\mathcal M^\cl$. Let $K$ and $K'$ be the dimensionality of $\mathcal M$ and $\mathcal M^\cl$, respectively, which satisfy $K'\leq K$. The dimensionality $K$ equals the number of linearly independent columns of $\mathsfit S$, i.e., the number of independent reactions, whereas $K'$ equals the number of linearly independent columns of $\mathsfit S_\cl$, i.e., the number of reactions that are independent even if we look at the closed species only. The difference $K-K'$ is nonzero when some of the independent reactions are dependent when we forget the open species.

\subsection{Reduced stoichiometric matrix}
\label{subsec_op_reduced}

As in the closed systems, we construct a reduced stoichiometric matrix $\mathsfit{A}\equiv (\Tns{A_i^\rho})_{i=1}^N {}_{\rho=1}^K$ as a basis of $\Im \mathsfit S$ and introduce the reduced reactions. In doing so, we impose additional conditions on the selection of basis of $\Im \mathsfit S$ to divide the reduced reactions into `change' reactions $\rho=1,\dots,K'$ and `emergent cycle' reactions $\rho=\hush{{K'}+1},\dots, K$.

Before discussing the division, we introduce related notation. We divide any vectors indexed by $\rho$ into the two parts. For example,
\begin{equation}
\begin{alignedat}{2}
    &\bm\eta_\chg \equiv (\eta_\rho)_{\rho=1}^{K'}, 
    &\quad &
    \bm\eta_\cyc \equiv (\eta_\rho)_{\rho=K'+1}^{K},\\
    &\bm\theta^\chg \equiv (\theta^\rho)_{\rho=1}^{K'},
    &\quad &
    \bm\theta^\cyc \equiv (\theta^\rho)_{\rho=K'+1}^{K},
\end{alignedat}
\end{equation}
where `$\chg$' stands for `change,' and `$\cyc$' stands for `cycle.' We also divide a reduced stoichiometric matrix $\mathsfit A \equiv (\Tns{A_i^\rho})_{i=1}^N {}_{\rho=1}^K$ into the four blocks
\begin{equation}
    \begin{alignedat}{2}
        &\Tns{{\mathsfit A}_\cl^\chg} \coloneqq (\Tns{A_i^\rho})_{i=1}^{N_\cl}\,{}_{\rho=1}^{K'},
        & \quad & \Tns{{\mathsfit A}_\cl^\cyc} \coloneqq (\Tns{A_i^\rho})_{i=1}^{N_\cl}\,{}_{\rho=K'+1}^K, \\
        &\Tns{{\mathsfit A}_\op^\chg} \coloneqq (\Tns{A_i^\rho})_{i=N_\cl+1}^{N}\,{}_{\rho=1}^{K'},
        & \quad & \Tns{{\mathsfit A}_\op^\cyc} \coloneqq (\Tns{A_i^\rho})_{i=N_\cl+1}^{N}\,{}_{\rho=K'+1}^K. 
    \end{alignedat}
    \label{op_S_divide_def}
\end{equation}
We also similarly divide the pseudo-inverse matrix $\mathsfit A^- \equiv ((A^-)\Tns{_\rho^i})$ into the four blocks, $\mathsfit A^-\Tns{_\chg^\cl}$, $\mathsfit A^-\Tns{_\chg^\op}$, $\mathsfit A^-\Tns{_\cyc^\cl}$, and $\mathsfit A^-\Tns{_\cyc^\op}$.

Now we can state the precise condition of the division as follows:
\begin{subequations}
\label{op_reducedcond_all}
\begin{align}
    &\text{The columns of $\Tns{{\mathsfit A}_\cl^\chg}$ form a basis of $\Im \mathsfit S_\cl$}, 
    \label{op_reducedcond1}\\
    &\Tns{{\mathsfit A}_\cl^\cyc} = 0.
    \label{op_reducedcond2}
\end{align}
\end{subequations}
The first condition~\eqref{op_reducedcond1} means that the `change' reduced reactions independently change the amounts of the closed species. The second condition~\eqref{op_reducedcond2} states that the `emergent cycle' reduced reactions only involve open species. In other words, the `emergent cycle' reduced reactions do not change the amounts of substances within the system.

To meet these conditions, we construct $\mathsfit{A}$ by the following procedure. For the first $K'$ columns, we collect $K'$ vectors from $\Im \mathsfit S$ that are linearly independent even when we only look at the closed species. For the remaining  $(\hush{K-{K'}})$ columns, we collect  $(\hush{K-{K'}})$ vectors of $\Im \mathsfit S$ such that all the $K$ vectors form a complete basis of $\Im \mathsfit S$. In doing so, we choose $(\hush{K-{K'}})$ vectors whose entries are nonzero only at the open species. This is always possible since any vectors in $\Im \mathsfit S$ are linearly dependent on the first $K'$ columns when we look at the closed species, and therefore we can always cancel out the nonzero entries at the closed species.

We also impose the following conditions on the choice of the pseudo-inverse matrix $\mathsfit A^-$: 
\begin{subequations}
\label{op_pinv_cond_all}
\begin{align}
& \mathsfit A^-\Tns{_\chg^\op} = 0 ,
\label{op_pinv_cond1}\\
& \text{$\mathsfit A^-\Tns{_\chg^\cl}$ is a pseudo-inverse matrix of $\Tns{{\mathsfit A}_\cl^\chg}$}.
\label{op_pinv_cond2}
\end{align}
\end{subequations}
The latter condition~\eqref{op_pinv_cond2} states that a part of $\mathsfit A^-$ is a pseudo-inverse matrix of the corresponding part of $\mathsfit A$, which is not trivial. These conditions can always be satisfied, as proven in Appendix~\ref{sec_appendix_pinv}.

To sum up, the reduced stoichiometric matrix $\mathsfit A$ and its pseudo inverse always have the form
\begin{equation}
    \mathsfit A = \mqty( 
        \Tns{{\mathsfit A}_\cl^\chg} & 0 \\ \Tns{{\mathsfit A}_\op^\chg} & \Tns{{\mathsfit A}_\op^\cyc}
    ), \quad 
    \mathsfit A^- = \mqty( 
        \mathsfit A^-\Tns{_\chg^\cl} &
        0 \\
        \mathsfit A^-\Tns{_\cyc^\cl} &
        \mathsfit A^-\Tns{_\cyc^\op}
    ).
    \label{op_S_divide}
\end{equation}

\vspace*{\baselineskip}
\textit{Example system 2}.---%
To demonstrate the dually flat geometry for open systems, we employ an enzymatic reaction system with species $(\X_1, \X_2, \X_3, \X_4, \X_5) = (\X_\sol, \E, \ES, \S, \P)$, respectively referring to the solvent, the enzyme, the enzyme--substrate complex, the substrate, and the product. We will substitute $(\sol,\E,\ES,\S,\P)$ for the suffixes $i=1,2,3,4,5$. We consider the substrate S and the product P to be the open species. We do not necessarily assume the ideal dilute property.

The system has two reactions:
\begin{equation}
\begin{alignedat}{2}
&\rho=1: &\quad& \E + \S \,\rightleftarrows\, \ES,\\
&\rho=2: &\quad& \ES\,\rightleftarrows\, \E + \P.
\end{alignedat}
\end{equation}
Since these two reactions are independent, the dimensionality of $\mathcal M$ is $K=2$. However, if we only look at the closed species, the first reaction transforms E into ES, and the second reaction transforms ES into E. They are redundant, and therefore, the dimensionality of $\mathcal M^\cl$ is $K'=1$.

We choose the reduced reactions to be
\begin{equation}
\begin{alignedat}{2}
&\rho=1: &\quad& \E + \S \,\rightleftarrows\, \ES,\\
&\rho=2: &\quad& \S\,\rightleftarrows\, \P.
\end{alignedat}   
\label{op_ex2_reduce}
\end{equation}
Then, the matrices are given by
\begin{equation}
\begin{aligned}
    & \mathsfit S = \mqty(0 & 0 \\ -1 & 1 \\ 1 & -1 \\ -1 & 0 \\ 0 & 1), \quad 
    \mathsfit A = 
    \left(
    \begin{array}{c|c}
        0 & 0 \\
        -1 & 0 \\
        1 & 0 \\
        \cline{1-2}
        -1 & -1 \\
        0 & 1 
    \end{array}
    \right), \quad
    \mathsfit c = \mqty( 1 & -1 \\ 0 & 1 ), \\
    & \mathsfit A^- = 
    \left(
    \begin{array}{ccc|cc}
        0 & 0 & 1 & 0 & 0 \\
        \cline{1-5}
        0 & 0 & 0 & 0 & 1 
    \end{array}
    \right).
\end{aligned}
\label{op_ex2_tS}
\end{equation}
The vertical and horizontal lines in $\mathsfit A$ and $\mathsfit A^-$ indicate the division of the matrices in Eq.~\eqref{op_S_divide}.

\subsection{Coordinates and the convex functions}
\label{subsec_op_coordinates}

\subsubsection{Construction}

We apply the construction for closed systems in the previous section to the total system. The \ueta-coordinates are defined through Eq.~\eqref{cl_defofeta}:
\begin{subequations}
\label{op_defofeta_all}
\begin{align}
    \bm n(\bm \eta) &= \bm n^\refr + \Tns{{\mathsfit A}_\cl^\chg} \bm \eta_\chg ,
    \label{op_defofeta1}\\
    \bm\nu (\bm\eta) 
    &= \bm\nu^\refr 
    + \Tns{{\mathsfit A}_\op^\chg}\bm\eta_\chg 
    + \Tns{{\mathsfit A}_\op^\cyc}\bm\eta_\cyc. 
    \label{op_defofeta2}
\end{align}
\end{subequations}
The convex function $\varphi(\bm\eta)$ is the total Gibbs free energy $G^\tot$ in Eq.~\eqref{ch_Gtot_approx} with the constant dropped. Although $\varphi(\bm\eta)$ is convex, it may not be strictly convex for two reasons. First, since $G^\tot$ linearly depends on $\bm\nu$, the inequality of convexity \eqref{ch_convexity} is saturated for two distributions $(\bm n, \bm\nu)$ and $(\bm n', \bm\nu')$  with $\bm n=\bm n'$ and $\bm\nu \neq \bm\nu'$. Second, since the mass of the molecules in the principal system is no longer conserved in open systems, two distributions $\bm n, \bm n'$ with $\bm n \propto \bm n'$ may reside in the same $\mathcal M^\cl(\bm n^\refr)$. For such a pair of distributions, the inequality of convexity~\eqref{ch_convexity} is saturated, and therefore $\varphi$ loses the strict convexity. Therefore, the following construction will follow the generalization discussed in Sec.~\ref{subsec_ge_generalization}\@. This difference makes the geometry of open systems unique.

The resulting quadruplet is as follows:
\begin{subequations}
\label{op_quadruplet_all}
\begin{empheq}[left={\empheqlbrace}\hspace{0.4em}]{alignat=2}
    &
    \begin{aligned}
        &\bm\eta_\chg(\bm n, \bm\nu) =  \mathsfit A^-\Tns{_\chg^\cl} ( \bm n - \bm n^\refr), \\
        &\bm\eta_\cyc(\bm n, \bm\nu) =  \mathsfit A^-\Tns{_\cyc^\cl} ( \bm n - \bm n^\refr) 
        +  \mathsfit A^-\Tns{_\cyc^\op} (\bm\nu -\bm\nu^\refr) ,
    \end{aligned}
    &\,\,\,\,&\Biggr\}
    \label{op_etaofn} \\
    &  \varphi(\bm\eta(\bm n, \bm\nu)) = 
     \bm\mu (\bm n) \cdot \bm n
    +  \bm\mu_\star \cdot \bm\nu , &&
    \label{op_varphi} \\
    &
    \begin{aligned}
        &\bm\theta^\chg (\bm n, \bm\nu) 
        =  \bm\mu (\bm n)\, \Tns{{\mathsfit A}_\cl^\chg}
        +  \bm\mu_\star \Tns{{\mathsfit A}_\op^\chg}, \\
        &\bm\theta^\cyc (\bm n, \bm\nu) =  \bm\mu_\star \Tns{{\mathsfit A}_\op^\cyc},
    \end{aligned}
    &&\Biggr\}
    \label{op_theta}
    \\
    &\psi(\bm\theta(\bm n, \bm\nu)) = -  \bm\mu (\bm n)\cdot \bm n^\refr -  \bm\mu_\star\cdot \bm\nu^\refr. &&
    \label{op_psi}
\end{empheq}
\end{subequations}
This quadruplet is almost similar to the quadruplet for closed systems \eqref{cl_quadruplet_all}, except that the chemical potentials of the open species are replaced by the constants $\bm\mu_\star$, and the amounts of the open species are by the total amounts $\bm\nu$. 

For later convenience, we perform an affine transformation \eqref{ge_affine2} that drops the constant term $-\bm\mu_\star \cdot \bm\nu^\refr$ from $\psi$ and adds the same constant to $\varphi$. The resulting convex functions, denoted by $\hat\varphi$ and $\hat\psi$, are
\begin{equation}
\hspace{-0.4em} \left\{ \hspace{0.4em}
\begin{aligned}
    & \hat\varphi(\bm\eta(\bm n, \bm\nu)) =\bm\mu(\bm n) \cdot \bm n + \bm\mu_\star \cdot \qty( \bm\nu - \bm\nu^\refr) ,
    \\
    & \hat\psi(\bm\theta(\bm n, \bm\nu))
    = - \bm\mu(\bm n) \cdot \bm n^\refr.
\end{aligned}
\right.
\label{op_quadruplet_trsf}
\end{equation}
We also rewrite $\hat\varphi$ explicitly as a function of $\bm\eta$:
\begin{align}
    \hat\varphi(\bm\eta_\chg, \bm\eta_\cyc) 
    & =  \bm\mu\qty(\strut\bm n(\bm\eta_\chg))\cdot \bm n(\bm\eta_\chg) \notag \\
    & \quad  + \bm\mu_\star \Tns{{\mathsfit A}_\op^\chg} \bm\eta_\chg + \bm\mu_\star \Tns{{\mathsfit A}_\op^\cyc} \bm\eta_\cyc,
    \label{op_varphi_trsf2} 
\end{align}
where we have used Eq.~\eqref{op_defofeta2}, and $\bm n(\bm\eta_\chg) $ is found in Eq.~\eqref{op_defofeta1}.

\subsubsection{Properties and interpretations}

\begin{figure}
    \centering
    \includegraphics[width=\hsize,clip,page=5]{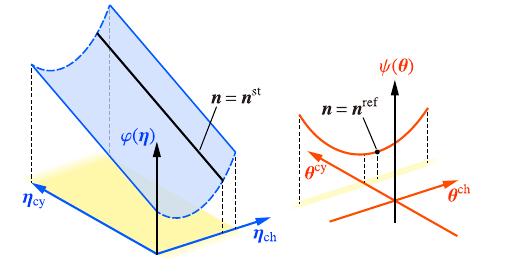}
    \caption{Schematics of the quadruplet of the dually flat geometry for open chemical reaction systems. The convex function $\varphi$ is the cumulative entropy production, i.e., the total Gibbs free energy of the system and the particle reservoirs together. Among the \ueta-coordinates, $\bm\eta_\cyc$ corresponds to the extent of `emergent cycle' reduced reactions. Since `emergent cycle' reduced reactions do not change the amounts of substances in the principal system, they do not change the affinity, i.e., the gradient of $\varphi(\bm\eta)$. Therefore, the gradient of $\varphi(\bm\eta)$ is independent of $\bm\eta_\cyc$, and $\varphi(\bm\eta)$ loses strict convexity. Since \utheta-coordinates are the gradient of $\varphi$, $\bm\theta^\cyc$ only takes a single value.}
    \label{fig_geometry_open}
\end{figure}

The followings are some observations on this quadruplet (see also Fig.~\ref{fig_geometry_open}). First, if we adopt the affine-transformed version $(\bm\eta, \hat\varphi(\bm\eta), \bm\theta,\hat\psi(\bm\theta))$, the total amounts do not appear alone but only appear as the difference $\bm\nu \nobreak-\nobreak \bm\nu^\refr$. Since the reference point is arbitrary, this means that the exact value of $\bm\nu$ is not relevant, and we only need their changes during a process.

The \ueta-coordinates are the extent of the reduced reactions and have a one-to-one correspondence with $(\bm n,\bm \nu)\in \mathcal M (\bm n^\refr, \bm\nu^\refr)$. Among the \ueta-coordinates,  $\bm\eta_\chg$ alone has a one-to-one correspondence with $\bm n \in \mathcal M^\cl(\bm n^\refr)$ [Eqs.~\eqref{op_defofeta1} and \eqref{op_etaofn}] and is not related to $\bm\nu$. The range of $\bm\eta_\chg$ is naturally determined by the positivity of the amounts of the closed species. In contrast, $\bm\eta_\cyc$ is related to the total amounts. Since the emergent cycles do not change the amounts of the closed species, the range of $\bm\eta_\cyc$ is not constrained by the positivity of the amounts of closed species. Therefore, we can consider the range of $\bm\eta_\cyc$ as extending infinitely.

If $K'<K$, the convex function $\hat\varphi$ is never strictly convex since it linearly depends on $\bm\eta_\cyc$ (see Eq.~\eqref{op_varphi_trsf2}). When we fix $\bm\eta_\cyc$, $\hat\varphi(\bm\eta_\chg,\bm\eta_\cyc)$ may or may not be a strictly convex function of $\bm\eta_\chg$. It loses the strict convexity if $\mathcal M^\cl(\bm n^\refr)$ accommodates two distributions $\bm n, \bm n'$ with $\bm n \propto \bm n'$. 

The \utheta-coordinates represent the negative of the affinities of the reduced reactions, as in closed systems. Since the strict convexity of $\varphi$ does not hold, the one-to-one correspondence between $\bm\eta$ and $\bm\theta$ is violated. In fact, the affinities of the emergent cycles $\bm\theta^\cyc$ take only a single value, $\bm\mu_\star \mathsfit A\Tns{_\op^\cyc} $.

The convex function $\hat\psi$ is introduced as a  function of $\bm\theta\equiv(\bm\theta^\chg,\bm\theta^\cyc)$, but it is also regarded as a function of $\bm\theta^\chg$ since $\bm\theta^\cyc$ only takes a single value.

We can interpret $\bm\theta$ and $\psi$ using the external field $\bm f$ as in closed systems. We introduce an external field $\bm f(\bm\theta)$ of the form $f^i(\bm\theta) = \sum_\rho \theta^\rho (A^-)\Tns{_\rho^i} + l^i(\bm\theta)$. Here, $(l^i)_{i=1}^N$ is a conservation law, i.e., a vector satisfying $\sum_i l^i\Tns{A_i^\rho} = 0$, and chosen so that $\bm f$ satisfies
\begin{equation}
    \bm f^\cl(\bm\theta) \cdot \bm n^\refr + 
    \bm f^\op(\bm\theta) \cdot \bm\nu^\refr 
    = 0,
    \label{op_v_cond}
\end{equation}
where $\bm f^\cl \equiv (f^i)_{i=1}^{N_\cl} $ and $\bm f^\op \equiv (f^i)_{N_\cl+1}^N$. The modified total Gibbs free energy is
\begin{equation}
    G^\tot(\bm n, \bm\nu;\, \bm f) = \qty[ \bm\mu(\bm n) - \bm f^\cl ] \cdot \bm n 
    +  \qty[ \vphantom{\bm f^\cl} \bm\mu_\star - \bm f^\op ] \cdot \bm\nu.
\end{equation}
We consider the minimization 
\begin{equation}
    \min_{(\bm n, \bm\nu) \,\in\, \mathcal M (\bm n^\refr, \bm\nu^\refr)} G^\tot(\bm n, \bm\nu;\, \bm f(\bm\theta)) .
    \label{op_variational}
\end{equation}
The minimizers are not unique, but any minimizer $(\bm n, \bm\nu)$ is a distribution corresponding to a single $\bm\theta$, confirming that the correspondence between $\bm\theta$ and $(\bm n, \bm\nu)$ is many-to-one. The potential $\psi(\bm\theta)$ is given by the minimum of Eq.~\eqref{op_variational}. We do not give the proof of these statements here since they are just rewrites of the results for closed systems. However, we would have some subtleties regarding the non-uniqueness of the minimizers. See Appendix \ref{sec_appendix_variational} for the justification of this point.

\vspace*{\baselineskip}
\textit{Example system 2}.---%
For our example system, we use the matrices in Eq.~\eqref{op_ex2_tS} and choose a reference point of the form $\bm n^\refr=(n_\sol^\refr, n_\E^\refr, 0)$ and $\bm\nu^\refr= (\nu_\S^\refr, \nu_\P^\refr)$, in which the enzyme--substrate complex is absent. Therefore, the quadruplet is
\begin{equation}
\left\{ \hspace{0.4em}
\begin{aligned}
    & \eta_1 = n_\ES, \quad \eta_2 = \nu_\P - \nu_\P^\refr, \\
    & \hat\varphi(\bm \eta) = \mu^\sol n_\sol + \mu^\E n_\E + \mu^\ES n_\ES  - \mu^\S_\star\eta_1 + (\mu^\P_\star - \mu^\S_\star)\eta_2, \\
    & \theta^1 = \mu^\ES - \mu^\E - \mu^\S_\star , \quad 
    \theta^2 = \mu^\P_\star - \mu^\S_\star,\\
    & \hat\psi(\bm\theta) = -\mu^\sol n_\sol^\refr - \mu^\E n_\E^\refr,
\end{aligned}
\right.
\end{equation}
where we have abbreviated the functional dependencies $\eta_\rho(\bm n, \bm\nu)$, $\theta^\rho(\bm n, \bm\nu)$, and $\mu^i(\bm n)$.

To obtain the inverted expression $\bm n(\bm\theta)$ and $\psi(\bm\theta)$, we introduce the external field $\bm f(\bm\theta)$ by
\begin{equation}
     \mqty(f^\sol(\bm\theta)\\ f^\E(\bm\theta)\\ f^\ES(\bm\theta)\\ f^\S(\bm\theta)\\
     f^\P(\bm\theta)) = \mqty(0\\0\\\theta^1\\0\\\theta^2) + \bm l(\bm\theta) = \mqty(0\\0\\\theta^1\\0\\\mu^\P_\star - \mu^\S_\star) + \bm l(\bm\theta),
\end{equation}
where $\bm l$ is a conservation law satisfying the constraint~\eqref{op_v_cond}
\begin{equation}
    f^\sol n_\sol^\refr + f^\E n_\E^\refr 
    + f^\S \nu_\S^\refr + f^\P \nu_\P^\refr= 0.
\end{equation}
For example, the choice
\begin{equation}
    l^1 = l^2 = 0, \quad l^3 = l^4 = l^5 = -\frac{\nu_\P^\refr}{\nu_\S^\refr+\nu_\P^\refr} (\mu^\P_\star - \mu^\S_\star)
\end{equation}
satisfy the constraint.
The modified total Gibbs free energy is calculated as
\begin{align}
    G^\tot(\bm n, \bm\nu; \bm f(\bm\theta)) \notag 
    &= \bm \mu(\bm n) \cdot \bm n - \theta^1  n_\ES + \mu_\star^\S(\nu_\S + \nu_\P)  \notag\\
    & \quad  -  l^3 (n_\ES + \nu_\S + \nu_\P).
\end{align}
The inverted relation $\bm n(\bm\theta)$ and $\psi(\bm\theta)$ are given by the minimization of this modified total Gibbs free energy. Since $n_\ES + \nu_\S + \nu_\P$ is conserved under the reactions, we have only to minimize the first line.

\subsection{Bregman divergence and differential geometry}
\label{subsec_op_divergence}

Although the convex function $\varphi$ violates the strict convexity, we can introduce the Bregman divergence and the line element using their definitions \eqref{ge_defofdiv2} and \eqref{ge_lineelement1}. The resulting Bregman divergence is
\begin{equation}
\D{ (\bm n,\bm\nu) || (\bm n',\bm\nu') } =  \qty[\bm\mu(\bm n) - \bm\mu(\bm n') ] \cdot \bm n.
\label{op_divergence}
\end{equation}
This form is identical to the closed case \eqref{cl_divergence}, noting that $\bm n$ denotes the amounts of the closed species in open systems. This divergence is determined solely by the amounts of the closed species and does not rely on the fictitious total amounts. To indicate this property clearly, we will hereafter write the divergence as 
\begin{equation}
    \D{ (\bm n,\bm\nu) || (\bm n',\bm\nu') } \equiv D(\bm n \| \bm n').
\end{equation}
Since the function $\varphi$ is not strictly convex, the divergence between two different points can be zero. For example, the divergence vanishes whenever $\bm n=\bm n'$, even if $\bm\nu\neq \bm\nu'$. In other words, the divergence is zero if the two points differ only by their $\bm\eta_\cyc$. The divergence also vanishes when $\bm n\propto\bm n'$.

The line element for open systems is also the same as that for closed systems~\eqref{cl_lineelement}, given by
\begin{equation}
ds^2 = d\bm \theta \cdot d\bm\eta =  d\bm\theta^\chg \cdot d\bm\eta_\chg  =d\bm\mu \cdot d\bm n,
\label{op_lineelement}
\end{equation}
where we have used $d\bm\theta^\cyc=0$ in the second equality since $\bm\theta^\cyc$ takes only a single value. The emergent cycles amount to zero length in this metric.

\vspace*{\baselineskip}
\textit{Ideal dilute solutions}.---%
For ideal dilute solutions, we can explicitly calculate the Bregman divergence and the line element using the expressions of $\bm\mu(\bm n)$~\eqref{ch_mu_ideal_all} and $\bm\pi(\bm n)$~\eqref{ch_pi_ideal_solute}. After some calculations, we arrive at
\begin{subequations}
\begin{align}
    &D(\bm n \| \bm n')
    = n_1RT \sum_{i=2}^{N_\cl} \qty( x_i \ln \frac{x_i}{x_i'} - x_i + x'_i),
    \label{op_divergence_ideal}  \\
    &ds^2 = n_1 RT \sum_{i=2}^{N_\cl} \frac{1}{x_i} (dx_i)^2,
\end{align}
\end{subequations}
where $x_i = n_i/n_1$ and $x'_i = n'_i/n_1$, and the sums are over the closed solutes. These are of the forms of a KL divergence and a generalized Fisher information. As in the closed systems, these expressions hold even if the amount of the solvent changes.

Although we have treated the solvent as a closed species, exchanging a small amount of solvent with the surroundings does not affect the divergence much. Recalling that the solvent is much more abundant than the solutes, we introduce $\varepsilon \ll 1$ as the typical value of the mole fractions of the solutes. Let us consider perturbing $n_1$ and $n'_1$ independently in the order of $n_1\varepsilon$, i.e., in the order of the amounts of the solutes. The perturbation alters $x_i$ and $x'_i$ in the order of $\varepsilon^2$ and, therefore, modifies $D(\bm n \| \bm n')$ in the order of $n_1\varepsilon^2$. However, this change is negligible compared to $D(\bm n \| \bm n')$ itself, which is of the order of $n_1\varepsilon$. Since all the geometrical quantities are derived from the divergence, the perturbation does not affect the geometry. Thus, we can allow a small exchange of the solvent without modifying our theory.

\subsection{Relation to the flows}
\label{subsec_op_flow}

The time evolution of the \ueta-coordinates can be related to the flows as in Eq.~\eqref{cl_etatoJ}. The relation provides an alternative route to introduce $\bm\eta$ and $\hat\varphi(\bm\eta)$ that does not rely on the fictitious concept of `total' amounts. Let us introduce $\bm\eta_\chg(t)$ by Eq.~\eqref{op_defofeta1} as a function of $\bm n(t)$. We next introduce $\bm\eta_\cyc(t)$ by integrating Eq.~\eqref{cl_etatoJ}:
\begin{equation}
    \eta_\rho(t) \coloneqq \sum_\sigma \Tns{c_\rho^\sigma} \! \int^t dt \, J_\sigma(t) \,\quad (\rho = K'+1,\dots, K).
    \label{op_eta_integral}
\end{equation}
The constant of integration is arbitrary, and this arbitrariness corresponds to the arbitrariness of $\bm\nu^\refr$ in the previous construction. Then, we can introduce $\hat\varphi(\bm\eta)$ by Eq.~\eqref{op_varphi_trsf2}. This way of introducing $\bm\eta$ and $\hat\varphi$ does not require $\bm\nu, \bm\nu^\refr$ and only relies on $\bm n(t)$, $\bm n^\refr$, $J_\rho(t)$, which are independent of any specific realizations of an open system. Therefore, our dually flat geometry for open systems applies to any realization of an open system.

The entropy production rate is expressed with the geometric quantities. Similarly to Eqs.~\eqref{cl_epr1} and \eqref{cl_epr2} for closed systems, we obtain two expressions of the entropy production rate:
\begin{subequations}
\begin{align}
    &T\dot\Sigma = - \frac{d\varphi}{dt} \geq 0,
    \label{op_epr1} \\
    &T \dot\Sigma = -\sum_{\rho} \theta^\rho \frac{d\eta_\rho}{dt} \geq 0.
    \label{op_epr2}
\end{align}
\end{subequations}
However, we cannot express $\varphi$ by a divergence, and therefore we cannot obtain an expression of entropy production rate with the divergence as in Eq.~\eqref{cl_epr3}. Instead, the divergence has an alternative role in open reaction systems, as discussed in the next section.

\section{Affine transformation and effective potential function}
\label{sec_steadystate}

In this section, we apply our dually flat geometry to the problem of finding an effective potential function for open systems. Consider a system that relaxes to a nonequilibrium steady state $\bm n^\st$, i.e., a state satisfying $dn_i/dt = \sum_\rho \Tns{S_i^\rho} \mathscr K_\rho(\bm n^\st) =0$ for $i=1,\dots,N_\cl$ and with a nonzero entropy production rate. If there are more than one such steady states, we focus on one of them. 

The relaxation to a nonequilibrium steady state in an open system can be compared to the relaxation to the equilibrium in a closed system. In closed systems, the Gibbs free energy as a function of $\bm n$ gives the cumulative entropy production, and its minimum corresponds to the equilibrium. In open systems, the cumulative entropy production cannot be written as a function of $\bm n$. Nevertheless, if we can construct an effective potential function of $\bm n$ for open systems that takes the minimum at the steady state, we can treat open and closed systems parallelly. Here we construct such a potential function that is natural from dually flat geometry.

\subsection{Construction}

\begin{figure}[t]
    \centering
    \includegraphics[width=\hsize,clip,page=6]{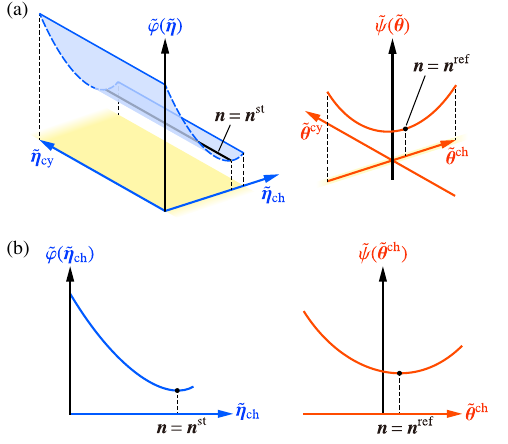}
    \caption{(a) Schematics of the affine-transformed quadruplet [Eq.~\eqref{po_quadruplet_all}]. Comparing this figure with Fig.~\ref{fig_geometry_open}, the minima of $\varphi$ are changed to the steady state, and the slopes of $\varphi$ along the $\tildebm\eta_\cyc$ axes are set to zero. Since the \utheta-coordinates are the gradient of $\varphi$, the single value of $\tildebm\theta^\cyc$ is moved to zero. (b) Schematics of the quadruplet $(\tildebm\eta_\chg,\tilde\varphi(\tildebm\eta_\chg),\tildebm\theta^\chg,\tilde\psi(\tildebm\theta^\chg))$, obtained by dropping $\tildebm\eta_\cyc$ and $\tildebm\theta^\cyc$ from the quadruplet in (a).}
    \label{fig_geometry_steadystate}
\end{figure}

Let us consider an affine transformation of the quadruplet $(\bm\eta,\varphi(\bm\eta),\bm\theta,\psi(\bm\theta))$ [Eq.~\eqref{op_quadruplet_all}] such that the new \utheta-coordinates vanish at the steady state. For this purpose, we take $\bm a \nobreak = \nobreak - \bm\theta(\bm n^\st)$, $\bm b=0$, $C = \psi(\bm\theta(\bm n^\st))$, and $\Tns{D_\rho^\sigma} = \Tns{\delta_\rho^\sigma}$ in the affine transformation [Eqs.~\eqref{ge_affine1} and \eqref{ge_affine2}], where we have used that $\bm\theta(\bm n,\bm \nu)$ is actually independent of $\bm\nu$ to write $\bm\theta(\bm n^\st)$. As a result, we obtain the following quadruplet $(\tildebm\eta,\tilde\varphi(\tildebm\eta),\tildebm\theta,\tilde\psi(\tildebm\theta))$ [Fig.~\ref{fig_geometry_steadystate}(a)]:
\begin{subequations}
\label{po_quadruplet_all}
\begin{empheq}[left=\hspace{2.2em}{\empheqlbrace}\hspace{0.4em}]{alignat=2}
    & \tilde{\bm\eta}(\bm n, \bm\nu) = \bm\eta(\bm n, \bm\nu), &&
    \label{po_quadruplet1}
    \\
    & \tilde\varphi(\tilde{\bm\eta}(\bm n, \bm\nu)) =\qty[ \bm\mu(\bm n)  -  \bm\mu (\bm n^\st) ] \cdot  \bm n, &&
    \label{po_quadruplet2}
    \\
    & 
    \begin{aligned}
        &\tildebm\theta^\chg(\bm n, \bm\nu) =  \qty[ \bm\mu(\bm n)  -  \bm\mu (\bm n^\st) ]  \mathsfit A\Tns{_\cl^\chg} \\
        &\tildebm\theta^\cyc(\bm n, \bm\nu) =  0, 
    \end{aligned}
    &\hspace{2.2em}&\Biggr\}
    \label{po_quadruplet3}
    \\
    & \tilde\psi(\tilde{\bm\theta}(\bm n, \bm\nu)) = 
    - \qty[ \bm\mu(\bm n)  -  \bm\mu (\bm n^\st) ] \cdot \bm n^\refr, &&
    \label{po_quadruplet4}
\end{empheq}
\end{subequations}
where $\bm\eta(\bm n, \bm\nu)$ is found in Eq.~\eqref{op_etaofn}. 

Our strategy is to use the thus constructed $\tilde\varphi$ as an effective potential function. We need to show that $\tilde\varphi$ is indeed a function of $\bm n$ and attains its minimum at $\bm n=\bm n^\st$. The former is obvious from the expression~\eqref{po_quadruplet2}. To prove the latter, we use $\partial \tilde\varphi/ \partial \tilde\eta_\rho = \tilde\theta^\rho = 0$ at the steady state, which implies that $\tilde\varphi$ takes its minimum at the steady state. Hereafter we use the abbreviation $\tilde\varphi(\bm n) \equiv \tilde\varphi(\tildebm\eta(\bm n,\bm\nu))$.

We can derive two suggestive expressions of the potential function $\tilde\varphi(\bm n)$. First, using the Euler equation~\eqref{ch_Euler}, we obtain
\begin{equation}
    \tilde\varphi(\bm n) = G(\bm n) - \sum_{i=1}^N \mu^i(\bm n^\st) n_i.
    \label{po_G_mun}
\end{equation}
This form is reminiscent of a Legendre transformation of the Gibbs free energy. Therefore, $\tilde\varphi$ can be considered a transformation of the Gibbs free energy. Second, we can exploit the relation between divergence and $\tilde\varphi$ in Eq.~\eqref{ge_div_varphionly} to obtain
\begin{equation}
    \tilde\varphi(\bm n) - \tilde\varphi( \bm n^\st) = D(\bm n \| \bm n^\st).
    \label{po_div}
\end{equation}
Here, this divergence can be considered as derived from the quadruplet for open systems~\eqref{op_quadruplet_all} or from the transformed quadruplet~\eqref{po_quadruplet_all}. They are identical since affine transformation does not change the divergence. In particular, if the system is an ideal dilute solution, we have the form of the KL divergence [Eq.~\eqref{op_divergence_ideal}]
\begin{equation}
  \tilde\varphi(\bm n) - \tilde\varphi( \bm n^\st) = n_1RT \sum_{i=2}^{N_\cl} \qty( x_i \ln \frac{x_i}{x_i^\st} - x_i + x_i^\st),
  \label{po_div_ideal}
\end{equation}
where $x_i^\st\coloneqq n_i^\st/n^\st_1$. This expression is similar to the Hatano--Sasa excess entropy production~\cite{hatano2001steady} or the boundary part of the nonadiabatic entropy~\cite{esposito2010three} in stochastic thermodynamics. Therefore, this expression suggests that our potential function $\tilde\varphi$ is the chemical-thermodynamic counterpart of the Hatano--Sasa excess entropy production. 

The same potential function for ideal dilute solutions has already appeared in some previous literature. Reference~\cite{rao2016nonequilibrium} introduced the potential function of the form \eqref{po_div_ideal} and pointed out that it is of the form of a KL divergence. Reference \cite{yoshimura2021information} introduced the potential function of the form~\eqref{po_G_mun} for ideal dilute solutions and also discussed that it is rewritten as a KL divergence. However, both these works are restricted to ideal dilute solutions. We here find an appropriate generalization for non-ideal solutions. We also find that $\tilde\varphi$ naturally arises from the affine transformation that sets the origin of the \utheta-coordinates to the steady state.

This potential function leads to a decomposition of entropy production rate into two components. The first component is the time derivative of the potential function. It is calculated similarly to Eq.~\eqref{op_epr2} from the affine-transformed quadruplet:
\begin{equation}
    -\frac{d\tilde\varphi}{dt} 
    = -\sum_\rho \tilde \theta^\rho \frac{d\tilde\eta_\rho }{dt} 
    = -\sum_\rho (\theta^\rho - \theta^\rho_\st) \frac{d\eta_\rho }{dt},
    \label{po_epr1} 
\end{equation}
where $\theta^\rho_\st \coloneqq \theta^\rho ( \bm n^\st)$, and the second equality is due to the affine transform. The other component is the difference between the original entropy production rate~\eqref{op_epr2} and the first component, calculated as
\begin{equation}
    T\dot\Sigma + \frac{d\tilde\varphi}{dt} = -\sum_\rho  \theta^\rho_\st \frac{d\eta_\rho }{dt}.
    \label{po_epr2}
\end{equation}
These two components give the decomposition of entropy production rate $T\dot\Sigma = - d\tilde\varphi/dt + (T\dot\Sigma + d\tilde\varphi/dt ) $. The first term is the component of entropy production rate that can be written as the time derivative of a potential function as in closed systems, while the second term is a component specific to open systems.

\subsection{Comparison with a closed system}

The effective potential function $\tilde\varphi$ allows us to compare the behaviors of an open system to those of a closed system. Let us consider a fictitious closed system whose species are $\X_1,\dots,\X_{N_\cl}$, whose Gibbs free energy is $\tilde\varphi(\bm n)$, and whose positive stoichiometric compatibility class is $\mathcal M^\cl(\bm n^\refr)$. The chemical potentials of this closed system are $\bm\mu(\bm n) - \bm\mu(\bm n^\st)$. 

First, we compare the dually flat geometry of the original open system with that of the fictitious closed system. The latter is constructed according to Sec.~\ref{sec_closed} as follows. We take $\tildebm\eta_\chg$ as \ueta-coordinates, which has the one-to-one correspondence with $\bm n\in\mathcal M^\cl(\bm n^\refr)$ [Eqs.~\eqref{op_defofeta1} and \eqref{op_etaofn}]. We take $\tilde\varphi(\bm n)$ as the convex function. Using these choices, we construct the \utheta-coordinates and the other convex function. The resulting quadruplet is given by $(\tildebm\eta_\chg,\tilde\varphi(\tildebm\eta_\chg),\tildebm\theta^\chg,\tilde\psi(\tildebm\theta^\chg))$. This quadruplet is the same as the one obtained by dropping $\tildebm\eta_\cyc$ and $\tildebm\theta^\cyc$ from the affine-transformed quadruplet $(\tildebm\eta,\tilde\varphi(\tildebm\eta),\tildebm\theta,\tilde\psi(\tildebm\theta))$ [Fig.~\ref{fig_geometry_steadystate}(b)]. Therefore, the dually flat geometry of the fictitious closed system is equivalent to that of the original open system up to an affine transformation and dropping the `emergent cycle' components.

From this equivalence, the fictitious closed system can be regarded as a `closed counterpart' of the open system with the same geometry. For example, if the original system is an ideal dilute solution, the closed counterpart is also an ideal dilute solution. Through this correspondence, any properties of closed systems derived through dually flat geometry can be readily rewritten as a property of open systems.

Next, we compare the time trajectory. In the closed counterpart, the trajectories should obey the second law of thermodynamics $d\tilde\varphi/dt < 0$ throughout the relaxation toward $\bm n^\st$. The question is whether the trajectories of the open system also satisfy $d\tilde\varphi/dt < 0$. If so, the relaxation processes of the open system cannot be distinguished from those of closed systems from the viewpoint of the thermodynamic potential. If not, the relaxation process of the open system is clearly discriminated from the relaxation processes of closed systems.

Whether $d\tilde\varphi/dt < 0$ holds or not strongly depends on the kinetics and the initial distribution. The detailed discussion is out of the scope of this paper, but here we state the following set of conditions as a sufficient condition for $d\tilde\varphi/dt<0$ for any initial distributions: (i) The kinetics satisfies the local detailed balance condition~\cite{avanzini2021nonequilibrium}. (ii) The kinetics is such that the rates of two reactions with the same reactants are proportional to each other. (iii) The steady state satisfies complex balancing~\cite{feinberg1972complex,feinberg2019foundations}. See Appendix~\ref{sec_appendix_CB} for the precise statements and the proof. Therefore, the relaxation processes of an open system satisfying (i)--(iii) are essentially similar to those of a closed system. Conversely, a system that breaks either one of these conditions may exhibit an increase in $\tilde\varphi$, which is a manifestation of relaxation behaviors specific to open systems.

\clearpage

\section{Discussion}
\label{sec_conclusion}

\subsection{Summary of the results}

We have introduced an information-geometric structure for chemical thermodynamics by explicitly constructing dual affine coordinates $\bm\eta$ and $\bm\theta$ with two convex functions $\varphi(\bm\eta)$ and $\psi(\bm\theta)$ [Eq.~\eqref{cl_quadruplet_all} for closed systems and Eq.~\eqref{op_quadruplet_all} for open systems]. Our main idea is to use linear combinations of the amounts of substances as an \ueta-coordinate system and take the cumulative entropy production as the convex function $\varphi(\bm\eta)$. As a result, we obtain \utheta-coordinates, equal to the affinities of the reduced reactions, and a new potential function $\psi(\bm\theta)$, interpreted as the equilibrium Gibbs free energy of a modified system. For open systems, the cumulative entropy production is not strictly convex, and therefore we have generalized dually flat geometry to accept such a convex function. 

We have identified the divergence and the metric of chemical thermodynamics. The metric is the Hessian of the cumulative entropy production [Eqs.~\eqref{cl_lineelement} and \eqref{op_lineelement}]. The divergence is the difference in chemical potentials multiplied by the amounts of substances in the first argument [Eqs.~\eqref{cl_divergence} and \eqref{op_divergence}]. These quantities reduce to the Fisher information and the KL divergence when the system is an ideal dilute solution, even if the amount of the solvent changes.

As an application, we have introduced an effective potential function $\tilde\varphi(\bm n)$ for the relaxation toward a steady state and decomposed the entropy production rate accordingly. The potential function has the form of a transform of the Gibbs free energy and is also rewritten as the divergence between an arbitrary state and the steady state [Eqs.~\eqref{po_G_mun} and \eqref{po_div}]. Since the potential function is derived from an affine transformation, it conserves the geometry induced from the original cumulative entropy production. Such a potential function would help compare an open system with a closed system, thereby highlighting their essential differences in their relaxation processes.

\subsection{Comparison with other geometric frameworks}

We briefly compare our dually flat geometry with other geometric frameworks for classical and chemical thermodynamics. Since our framework derives from the information geometry of stochastic thermodynamics, other frameworks discussed here are not the direct antecedents of our work. Nevertheless, our geometry has some similarities and relationships with them. Note that we are not to claim that our geometry is more suitable than other frameworks. Each geometry has its own scope of application and advantages. 
 
One of such frameworks is the Riemannian geometry of classical thermodynamics~\cite{weinhold1975metric,ruppeiner1979thermodynamics,salamon1983length}, dating back to the 1970s. This framework has been used to discuss, for example, fluctuations in equilibrium thermodynamics and the characterization of interactions among component particles~\cite{rupperiner1995revmodphys}. This framework takes the Hessian of a thermodynamic potential as a metric. While a metric is a local property, it also induces some global properties, such as the length between two distant points by the integral of the metric. Our dually flat geometry has the same metric [Eq.~\eqref{cl_lineelement}] but adopts a different global structure characterized by the divergence [Eq.~\eqref{cl_divergence}], which is an asymmetric measure of how distant the two points are. The divergence is more closely related to thermodynamic quantities and, therefore, a useful tool to decompose thermodynamic quantities as in Sec.~\ref{sec_steadystate}\@. This global structure reduces to the local metric for infinitesimally close points [Eq.~\eqref{ge_lineelement_all}]. In this sense, our dually flat geometry provides modern concepts such as dual affine coordinates and the divergence consistent with the metric in those historical works. 

Another framework is the contact geometry of classical thermodynamics~\cite{hermann1973geometry, mrugala1978geometrical,mrugala1991contact}. Contact geometry involves a function called contact Hamiltonian, which incurs the Hamiltonian dynamics. In Ref.~\cite{grmela2012fluctuations}, the mass action kinetics is reproduced as a Hamiltonian dynamics, and some extensions such as the inclusion of inertial terms are discussed. The contact geometry for classical thermodynamics treats intensive and extensive variables as independent variables. The actual correspondence of these variables is achieved only on a special subspace called the Legendre manifold. Therefore, the contact geometry mainly investigates the Legendre manifold \textit{from outside} of it with its relation to kinetics, such as a kinetics that relaxes toward the Legendre manifold. In contrast, our dually flat geometry is formulated \textit{within} the Legendre manifold (if we barrow the terminology of contact geometry) and suitable for investigating the structure within the Legendre manifold using the dual affine coordinates and the divergence. In this regard, our dually flat geometry is complementary to contact geometry.

The Riemannian geometry and the contact geometry have been combined to form a framework called geometrothermodynamics~\cite{quevedo2007geometro,vazquez2010}, using the Legendre invariance of the metric as the guiding principle. This framework has also been applied to chemical reaction systems~\cite{quevedo2014geometric}, finding the singularity of the metric at the equilibrium. This framework and our results are compared similarly to the above two comparisons.

\subsection{Conclusion and outlook}

We have developed an information geometric structure for chemical reaction systems, including open and non-ideal systems. Our construction of information geometry will be a useful tool for the nonequilibrium thermodynamics of chemical reaction systems, particularly for finding decompositions of quantities and new inequalities, as demonstrated in Sec.~\ref{sec_steadystate}.

In investigating such applications further, one can learn from existing results in stochastic thermodynamics. Since information geometry has been used in stochastic thermodynamics to obtain many results, our construction would enable us to import those results into chemical thermodynamics. In other words, our construction serves as a new structural bridge between chemical and stochastic thermodynamics.

Aside from these applications within chemical thermodynamics, our construction also provides a new methodology for information geometry of thermodynamics in general. First, the information geometry derived from a potential function violating the strict convexity has not been considered even in stochastic thermodynamics. Such an information geometry would be helpful in studying non-detailed balanced systems in stochastic thermodynamics. Second, this work suggests a strategy to find information-geometric structure in other thermodynamic frameworks than stochastic and chemical thermodynamics, such as the thermodynamic of fluid systems~\cite{fitts1962nonequilibrium}. The lesson is to take a proper linear combination of some natural variables as the coordinates and take the cumulative entropy production as the convex function. With this strategy, we may be able to bridge more thermodynamic frameworks using information geometry.

\vspace*{2\baselineskip}

\noindent \textit{Note added}: Recently, a Hessian geometric structure for chemical thermodynamics was constructed by Y. Sughiyama, D. Loutchko, A. Kamimura, and T. J. Kobayashi in their two successive papers~\cite{sughiyama2021, kobayashi2021}. Since Hessian geometry and dually flat geometry share several concepts such as dual coordinates and Bregman divergence, they also discuss these concepts in chemical thermodynamics. One of the two papers~\cite{sughiyama2021} appeared on arXiv slightly before our first submission of this paper to arXiv, while the other~\cite{kobayashi2021} appeared slightly after our submission. Still, we had independently conducted our study and had previously made an oral presentation of our result at the Physical Society of Japan 2021 Autumn Meeting.

\acknowledgements

We thank Kohei Yoshimura for fruitful discussions on chemical thermodynamics. S. I. is supported by JSPS KAKENHI Grants No.\ 19H05796, No.\ 21H01560, and No.\ 22H01141, JST Presto Grant No.\ JPMJPR18M2, and UTEC-UTokyo FSI Research Grant Program.

\appendix

\section{Approximation of the Gibbs free energy for open systems}
\label{sec_appendix_Gtot}

In this section, we derive the approximation of total Gibbs free energy in Eq.~\eqref{ch_Gtot_approx}. First, we introduce quantities of the reservoirs without any approximation. The amounts of the open species in the reservoirs are the difference between the total amounts and the amounts in the system:
\begin{equation}
    \bm\pi^\res \coloneqq  \bm\nu - \bm\pi.
\end{equation}
For simplicity, we assume that each open species is held in only one of the reservoirs. Therefore, the sum of the Gibbs free energies of the reservoirs is regarded as a function of $\bm\pi^\res$, denoted by $G^\res(\bm\pi^\res)$. We also define the chemical potentials of the reservoirs by
\begin{equation}
    \mu_\res^i(\bm\pi^\res) \coloneqq \frac{\partial G^\res}{\partial \pi^\res_i} \qquad (\text{$i$: open species}),
\end{equation}
and write $\bm \mu_\res \equiv (\mu_\res^i)_{i=N_\cl+1}^N$. An infinitesimal change $d\bm\pi^\res$ induces the change in $G^\res$ as
\begin{equation}
    dG^\res = \bm\mu_\res(\bm\pi^\res) \cdot d \bm\pi^\res.
    \label{ap_infinitesimal}
\end{equation}

Now we approximate the Gibbs free energy of the reservoirs, while we do not approximate that of the principal system. Since the reservoirs are vast, we can approximate that the chemical potentials of the reservoirs are fixed at $\bm\mu_\star$ during a process. With this approximation, we integrate Eq.~\eqref{ap_infinitesimal} to obtain
\begin{align}
G^\res(\bm\pi^\res) - G^\res(\bm\pi^{\res,\refr})
\simeq \bm\mu_\star \cdot \qty[\bm\pi^\res - \bm\pi^{\res,\refr} ],
\label{ap_Gres_approx}
\end{align}
where $\bm\pi^{\res,\refr}$ is an arbitrary reference point. Inserting this approximation, the total Gibbs free energy $G^\tot$ is approximated as
\begin{align}
    G^\tot(\bm n, \bm\nu)
    & = G(\bm n, \bm \pi(\bm n)) + G^\res(\bm\pi^\res) \notag \\
    &  \simeq \bm\mu(\bm n, \bm\pi(\bm n)) \cdot \bm n 
    +  \bm\mu_\star \cdot \bm \pi(\bm n) \notag \\  \nopagebreak
    &  \quad + G^\res(\bm\pi^{\res,\refr}) 
    + \bm\mu_\star \cdot \qty[\qty(\bm\nu - \bm\pi(\bm n)) - \bm\pi^{\res,\refr}  ] \notag \\
    & = \bm\mu(\bm n, \bm\pi(\bm n))\cdot \bm n 
    + \bm\mu_\star \cdot \bm\nu 
    + \mathrm{const.},
    \label{ap_Gtot_approx}
\end{align}
where the constant does not depend on $(\bm n,\bm\nu)$. This expression reproduces the approximation in the main text~\eqref{ch_Gtot_approx}.

\section{Properties of the pseudo-inverse matrices of a reduced stoichiometric matrix}
\label{sec_appendix_pinv}

In this section, we discuss some properties of the pseudo-inverse matrices of a reduced stoichiometric matrix $\mathsfit A$. A reduced stoichiometric matrix is an $N\times K$ matrix ($N > K$) of rank $K$. Its pseudo-inverse matrix $\mathsfit A^-$ is defined by any matrix satisfying
\begin{equation}
    \sum_{\rho,j}\Tns{A_i^\rho}\Tns{{(A^-)}_\rho^j}\Tns{A_j^\sigma}
    = \Tns{A_i^\sigma}.
    \label{ps_pseudoinv}
\end{equation}
Since any rectangle matrix has a pseudo-inverse matrix, so does $\mathsfit A$. However, pseudo-inverse matrices are not unique. Let $\mathsfit A^-$ be any one of them.
Then, we can prove the following relations:
\begin{subequations}
\begin{align}
    & \sum_i (A^-)\Tns{_\rho^i} \Tns{A_i^\sigma} = \Tns{\delta_\rho^\sigma} 
    \label{ps_pinv1}\\
    & \sum_{\rho,j} \Tns{A_i^\rho} (A^-)\Tns{_\rho^j} (n_j - n'_j) = n_i - n'_i,
    \label{ps_pinv2}
\end{align}
\end{subequations}
where $\bm n$ and $\bm n'$ are arbitrary two distributions in one positive stoichiometric compatibility class. To prove Eq.~\eqref{ps_pinv1}, we rewrite the definition \eqref{ps_pseudoinv} as $\sum_{\rho,j} \Tns{A_i^\rho}\qty[\Tns{{(A^-)}_\rho^j} \Tns{A_j^\sigma} - \Tns{\delta_\rho^\sigma}]= 0$. Since $\Tns{A_i^\rho}$ is of rank $K$, the kernel of $\mathsfit A$ is the null space. Therefore, we obtain $\qty[ \sum_j\Tns{{(A^-)}_\rho^j} \Tns{A_j^\sigma} - \Tns{\delta_\rho^\sigma}]= 0$, which is equivalent to Eq.~\eqref{ps_pinv1}. To prove Eq.~\eqref{ps_pinv2}, we recall that the difference between two points in one positive stoichiometric compatibility class is a linear combination of $(\Tns{A_i^\rho})_{i=1}^N$. Therefore, we can always write as $n_j - n_j' = \sum_\sigma \Tns{A_j^\sigma} \xi_\sigma$ using some constants $(\xi_\sigma)_{\sigma=1}^K$. Inserting this expression into the left-hand side of Eq.~\eqref{ps_pinv2} and using the definition \eqref{ps_pseudoinv}, we easily obtain Eq.~\eqref{ps_pinv2}.

Next, we prove the following fact used in Sec.~\ref{subsec_op_reduced}\@. By dividing the reduced stoichiometric matrix and its pseudo-inverse matrix as
\begin{equation}
    \mathsfit A = \mqty( 
        \mathsfit A\Tns{_\cl^\chg} & \mathsfit A\Tns{_\cl^\cyc} \\ \mathsfit A\Tns{_\op^\chg} & \mathsfit A\Tns{_\op^\cyc}
    ), \quad
    \mathsfit A^- = \mqty( 
        \mathsfit A^-\Tns{_\chg^\cl} &
        \mathsfit A^-\Tns{_\chg^\op} \\
        \mathsfit A^-\Tns{_\cyc^\cl} &
        \mathsfit A^-\Tns{_\cyc^\op}
    ),
\end{equation}
we can choose a pseudo-inverse matrix $\mathsfit A^-$ such that 
\begin{subequations}
\label{ps_pinv_cond_all}
\begin{align}
& \mathsfit A^-\Tns{_\chg^\op} = 0 ,
\label{ps_pinv_cond1} \\
& \text{$\mathsfit A^-\Tns{_\chg^\cl}$ is a pseudo-inverse of $\Tns{{\mathsfit A}_\cl^\chg}$},
\label{ps_pinv_cond2} \\
& \text{$\mathsfit A^-\Tns{_\cyc^\op}$ is a pseudo-inverse of $\Tns{{\mathsfit A}_\op^\cyc}$}.
\label{ps_pinv_cond3}
\end{align}
\end{subequations}
To prove this fact, we recall the following properties of $\mathsfit A$,
\begin{subequations}
\label{ps_tS_cond_all}
\begin{align}
& \mathsfit A\Tns{_\cl^\cyc} = 0,
\label{ps_tS_cond1} \\
&\text{The columns of $\mathsfit A\Tns{_\cl^\chg}$ are linearly independent}, 
\label{ps_tS_cond2} \\
&\text{The columns of $\mathsfit A\Tns{_\op^\cyc}$ are linearly independent}, 
\label{ps_tS_cond3}
\end{align}
\end{subequations}
which follow from the construction of $\mathsfit A$ in Sec.~\ref{subsec_op_reduced}\@. Exploiting these properties, we can construct $\mathsfit A^-$ satisfying the conditions~\eqref{ps_pinv_cond_all} as follows. Let $\mathsfit R\Tns{_\chg^\cl}$ be any pseudo-inverse matrix of $\mathsfit A\Tns{_\cl^\chg}$, and $\mathsfit R\Tns{_\cyc^\op}$ be any pseudo-inverse matrix of $\mathsfit A\Tns{_\op^\cyc}$. These matrices satisfy
\begin{equation}
    \mathsfit R\Tns{_\chg^\cl} \mathsfit A \Tns{_\cl^\chg} = \mathsfit I_\chg, \quad 
    \mathsfit R\Tns{_\cyc^\op} \mathsfit A \Tns{_\op^\cyc} = \mathsfit I_\cyc,
    \label{ps_partialpinv_simple}
\end{equation}
where $\mathsfit I_\chg$ and $\mathsfit I_\cyc$ are the identity matrices of size $K'\times K'$ and $(\hush{K-K}')\times( \hush{K-K}')$, respectively. Equation~\eqref{ps_partialpinv_simple} can be proven similarly to Eq.~\eqref{ps_pinv1} thanks to the properties \eqref{ps_tS_cond2} and \eqref{ps_tS_cond3}. Then, the matrix 
\begin{equation}
    \mathsfit A^- \coloneqq \mqty(\mathsfit R\Tns{_\chg^\cl} & 0 \\ - \mathsfit R\Tns{_\cyc^\op} \mathsfit A \Tns{_\op^\chg} \mathsfit R\Tns{_\chg^\cl} & \mathsfit R\Tns{_\cyc^\op})
    \label{ps_defofR}
\end{equation}
is a desired pseudo-inverse matrix. Indeed, the matrix product $\mathsfit A^- \mathsfit A$ is calculated as
\begin{align}
    \mathsfit A^- \mathsfit A 
    &= \mqty(
        \mathsfit R\Tns{_\chg^\cl} & 0 \\ 
        - \mathsfit R\Tns{_\cyc^\op} \mathsfit A \Tns{_\op^\chg} \mathsfit R\Tns{_\chg^\cl} &
        \mathsfit R\Tns{_\cyc^\op})
    \mqty( 
        \mathsfit A\Tns{_\cl^\chg} & 0 \\
        \mathsfit A\Tns{_\op^\chg} &
        \mathsfit A\Tns{_\op^\cyc}
    )  \notag \\
    &= \mqty(
        \mathsfit R\Tns{_\chg^\cl} \mathsfit A\Tns{_\cl^\chg}\,\,  &
        0 \\
        - \mathsfit R\Tns{_\cyc^\op} \mathsfit A \Tns{_\op^\chg} \mathsfit R\Tns{_\chg^\cl} \mathsfit A\Tns{_\cl^\chg} +
        \mathsfit R\Tns{_\cyc^\op}
        \mathsfit A\Tns{_\op^\chg}\,\, &
        \mathsfit R\Tns{_\cyc^\op}
        \mathsfit A\Tns{_\op^\cyc}
    ) \notag \\
    &= \mqty(
        \mathsfit I_\chg  & 0 \\
        0 & \mathsfit I_\cyc 
    ),
\end{align}
where we have used Eq.~\eqref{ps_partialpinv_simple} in the last equality. Therefore, the definition of pseudo-inverse matrix  $\mathsfit A^- \mathsfit A  \mathsfit A^- =  \mathsfit A$ is satisfied. The first condition \eqref{ps_pinv_cond1} is obviously satisfied from the construction \eqref{ps_defofR}. The second~\eqref{ps_pinv_cond2} and the third condition~\eqref{ps_pinv_cond3} are also met from the definition of $\mathsfit R\Tns{_\chg^\cl}$ and $\mathsfit R\Tns{_\cyc^\op} $.

\section{Variational expression of \texorpdfstring{$\bm n(\bm\theta)$ and $\psi(\bm\theta)$}{n and psi}}
\label{sec_appendix_variational}

Our goal in this section is to derive the expression of $\bm n(\bm\theta)$ and $\psi(\bm\theta)$ for closed systems [Eqs.~\eqref{cl_noftheta} and  \eqref{cl_psioftheta}] and those for open systems [Eq.~\eqref{op_variational}] from the variational formula of Legendre duality. The variational formula of Legendre duality is the following set of expressions (see, for example, Refs.~\cite{rockafellar1996,tasaki2022}):
\begin{subequations}
\label{vr_variational1}
\begin{gather}
    \bm \eta(\bm\theta) = \argmax_{\bm\eta} \qty[ \bm\eta\cdot \bm\theta - \varphi(\bm\eta) ], 
    \label{vr_variational1a}\\
    \psi(\bm\theta) = \max_{\bm\eta} \qty[ \bm\eta\cdot \bm\theta - \varphi(\bm\eta) ] ,
    \label{vr_variational1b}
\end{gather}
\end{subequations}
and 
\begin{subequations}
\label{vr_variational2}
\begin{gather}
    \bm \theta(\bm\eta) = \argmax_{\bm\theta} \qty[ \bm\theta\cdot \bm\eta - \psi(\bm\theta) ], 
    \label{vr_variational2a}\\
    \varphi(\bm\eta) = \max_{\bm\theta} \qty[ \bm\theta\cdot \bm\eta - \psi(\bm\theta) ].
    \label{vr_variational2b}
\end{gather}
\end{subequations}
These expressions are even valid when $\varphi$ is not strictly convex nor twice differentiable. If the correspondence between $\bm\eta$ and $\bm\theta$ is not one-to-one, the maximizers in these maximization problems may not be unique. In this case, we should understand Eq.~\eqref{vr_variational1a} as the equality between the set of the maximizers of the right-hand side and the set of all $\bm\eta$ corresponding to a single $\bm\theta$. Equation~\eqref{vr_variational2a} should also be understood similarly.

We first derive the formula for closed systems by rewriting the variational formula of Legendre transformation \eqref{vr_variational1}. First, since $\bm\eta$ and $\bm n\in\mathcal M(\bm n^\refr)$ admit the one-to-one correspondence, we obtain 
\begin{subequations}
\label{vr_variation2_all}
\begin{align}
&\bm n(\bm\theta) = \argmax_{\bm n\,\in\,\mathcal M(\bm n^\refr)}\qty[\strut \bm\theta \cdot \bm\eta(\bm n)  - \varphi(\bm\eta(\bm n))], 
\label{vr_variation2_a} \\
&\psi (\bm\theta) = \max_{\bm n\,\in\,\mathcal M(\bm n^\refr)}\qty[\strut \bm\theta \cdot \bm\eta(\bm n) - \varphi(\bm\eta(\bm n))]
\label{vr_variation2_b} .
\end{align}
\end{subequations}
Note that $\bm\theta$ and $\bm n$ are treated as independent of each other within the square brackets. Next, we rewrite the maximand by using the definition of the external field $\bm f(\bm\theta)$ [Eqs.~\eqref{cl_thetaofv} and \eqref{cl_v_cond}] as follows:
\begin{align}
    &\bm\theta \cdot \bm\eta(\bm n) - \varphi(\bm\eta(\bm n)) \notag \\
    &= \sum_{j,\rho,i}  f^j \Tns{A_j^\rho} (A^-)\Tns{_\rho^i} (n_i - n_i^\refr) -\sum_i \mu^i(\bm n) n_i \notag \\
    &= \sum_{i}  f^i (n_i - n_i^\refr) -\sum_i \mu^i(\bm n) n_i \notag \\
    &= - G(\bm n; \bm f(\bm\theta)),
\end{align}
where we have used the property of $\mathsfit A$ in Eq.~\eqref{ps_pinv2} in the second equality. Inserting this result into Eq.~\eqref{vr_variation2_all}, we reproduce the desired results [Eqs.~\eqref{cl_noftheta} and  \eqref{cl_psioftheta}] for closed systems.

The similar result for open systems directly follows from this discussion for closed systems by replacing the Gibbs free energy $G$ with $G^\tot$. The only difference is that $G^\tot$ is no longer strictly convex, and therefore the correspondence between $\bm\eta$ and $\bm\theta$ is many-to-one. Nevertheless, the variational formula of Legendre duality \eqref{vr_variational1}--\eqref{vr_variational2} is valid for such cases, and therefore the above derivation for closed systems readily applies to open systems.

\section{A sufficient condition for the monotonicity of the effective potential function}
\label{sec_appendix_CB}

In this section, we state and prove a sufficient condition for the monotonic decrease of the potential function $\tilde\varphi$, introduced in Sec.~\ref{sec_steadystate}\@. The following is based on a similar discussion for ideal dilute solutions with mass action kinetics in Refs.~\cite{rao2016nonequilibrium,ge2016nonequilibrium}.

We first introduce the necessary notation. In the reactions in Eq.~\eqref{ch_reaction}, we call the reaction converting the left-hand side into the right-hand side \textit{the $\rho$th forward reaction} and the reaction in the opposite direction \textit{the $\rho$th backward reaction}. We write the rate of the $\rho$th forward (backward) reaction as $K_\rho(\bm n)$ ($K_{-\rho}(\bm n)$)\footnote{The symbol $K_\rho(\bm n)$ should not be confused with the symbol $K$ used in the main text to denote the dimensionality of spaces.}. These rates satisfy 
\begin{equation}
    \mathscr K_\rho(\bm n)
    = K_\rho(\bm n) - K_{-\rho}(\bm n),
\end{equation}
where $\mathscr K_\rho(\bm n)$ has been introduced in Eq.~\eqref{ch_kinetics}. We also define the stoichiometric coefficients $\Tns{\lambda_i^\rho}$ and the affinity $F^\rho(\bm n)$ for $\rho=-1,\dots,-M$ for notational simplicity:
\begin{equation}
    \Tns{\lambda_i^{-\rho}} \coloneqq \Tns{\kappa_i^\rho}, \quad 
    F^{-\rho}(\bm n) \coloneqq -F^\rho(\bm n), \quad (\rho=1,\dots,M).
\end{equation}
In the remainder of this appendix, indices $\rho$ and $\sigma$ run over $\pm 1,\pm 2,\dots, \pm M$ unless otherwise noted. We also introduce the set of all the vectors appearing as the stoichiometric coefficients:
\begin{equation}
    \mathcal C \coloneqq \Set{ \bm\lambda^\rho | \rho=\pm 1 ,\dots, \pm M },
\end{equation}
where $\bm\lambda^\rho \equiv (\Tns{\lambda_i^\rho})_{i=1}^N$.

Now we state the three conditions. The first condition is the local detailed balance condition~\cite{avanzini2021nonequilibrium}, defined as follows:
\begin{equation}
    RT \ln \frac{K_\rho(\bm n)}{K_{-\rho}(\bm n)} = 
    F^\rho(\bm n).
\end{equation}
This condition connects the kinetics and the thermodynamic quantities. It is a sufficient condition for the second law of thermodynamics and is often assumed in non-ideal chemical reaction systems (for example, Refs.~\cite{avanzini2021nonequilibrium,yoshimura2021prl}).

The second condition is the proportionality of the rates departing from the same $\bm\lambda^\rho$. For any pair $\rho$ and $\sigma$ satisfying $\bm\lambda^\rho = \bm\lambda^\sigma$, we require that
\begin{equation}
    \text{$\displaystyle\frac{K_\rho(\bm n)}{K_\sigma(\bm n)}$ is independent of $\bm n$}.
    \label{cb_regularity}
\end{equation}
This condition is easily interpreted when the reaction rates are the product of the collision rate of the reactant molecules and the probability that the reaction occurs after a collision. The collision rates are equal for two reactions with the same set of reactants $\bm\lambda^\rho$. Therefore, the ratio in Eq.~\eqref{cb_regularity} reduces to the ratio of the probability that the reaction occurs. If the probability is independent of the surrounding amounts of substances $\bm n$, the condition~\eqref{cb_regularity} holds.

The third condition is a property of the steady state called the \textit{complex balancing}~\cite{feinberg1972complex,feinberg2019foundations}. The steady state is said to be complex balanced if, for any $\bm y\in\mathcal C$,
\begin{equation}
    \sum_{\rho\,|\, \bm\lambda^\rho = \bm y}  K_\rho(\bm n^\st)  
    =
    \sum_{\rho\,|\, \bm\lambda^\rho = \bm y}  K_{-\rho}(\bm n^\st).
\end{equation}
In other words, the total flow entering $\bm y$ equals the total flow exiting from $\bm y$. This condition is stronger than the steady-state condition $\sum_\rho \mathscr K_\rho(\bm n^\st) \Tns{S_i^\rho} =0$ ($i$:~closed species) in general. However, every steady state is complex balanced for a special class of systems called deficiency zero networks~\cite{feinberg2019foundations}.

We remark that, for ideal dilute solutions with mass action kinetics~\cite{lund1965guldberg,horn1972general,feinberg2019foundations}, the first and the second conditions are always satisfied. Therefore, the only condition for $d\tilde\varphi/dt\leq 0$ is the complex balancing. This result is known in Refs.~\cite{rao2016nonequilibrium,ge2016nonequilibrium}.

With these conditions, we prove the non-positivity of $d\tilde\varphi/dt$ as follows. First, we rewrite $d\tilde\varphi/dt$ in Eq.~\eqref{po_epr1} with the aid of Eqs.~\eqref{cl_F_theta} and \eqref{cl_etatoJ}, which are valid also for open systems:
\begin{align}
    \frac{d\tilde\varphi}{dt}
    &= - \sum_{\rho=1}^M \mathscr K_\rho(\bm n) \qty[ F^\rho(\bm n) - F^\rho(\bm n^\st) ] \notag  \\
    &= - \hspace{-0.5em} \sum_{\rho=\pm 1,\dots,\pm M}  K_\rho(\bm n) \qty[ F^\rho(\bm n) - F^\rho(\bm n^\st) ] .
\end{align}
Next, we use the first condition and $\ln x \leq x - 1$ for $x>0$ to obtain
\begin{align}
    \frac{1}{RT}\frac{d\tilde\varphi}{dt}
    &=  \sum_\rho K_\rho(\bm n) \ln\qty[ \frac{K_{-\rho}(\bm n)K_\rho(\bm n^\st)}{K_\rho(\bm n)K_{-\rho}(\bm n^\st)} ] \notag \\
    &\leq  \sum_\rho K_\rho(\bm n) \qty[ \frac{K_{-\rho}(\bm n)K_\rho(\bm n^\st)}{K_\rho(\bm n)K_{-\rho}(\bm n^\st)} -1 ] \notag \\
    &= \sum_\rho  K_\rho(\bm n^\st) \qty[ \frac{K_{-\rho}(\bm n)}{K_{-\rho}(\bm n^\st)} - \frac{ K_\rho(\bm n)}{ K_\rho(\bm n^\st)}] \notag \\
    &= - \sum_\rho  \qty[K_\rho(\bm n^\st) - K_{-\rho}(\bm n^\st) ]  \frac{ K_\rho(\bm n)}{ K_\rho(\bm n^\st)} \notag \\
    &= - \sum_{\bm y \in \mathcal C} \sum_{\rho \,|\, \bm\lambda^\rho = \bm y} \qty[K_\rho(\bm n^\st) - K_{-\rho}(\bm n^\st) ]  \frac{ K_\rho(\bm n)}{ K_\rho(\bm n^\st)},
\end{align}
where $\sum_\rho$ is the sum over $\rho=\pm 1,\dots,\pm M$. It suffices to show that the last line is zero. To show this, we rearrange the second condition to obtain ${K_\rho(\bm n)}/{K_\rho(\bm n^\st)} = {K_\sigma(\bm n)}/{K_\sigma(\bm n^\st)} $ if $\bm\lambda^\rho = \bm\lambda^\sigma$. Therefore, we can move $K_\rho (\bm n)/K_\rho(\bm n^\st)$ in front of the second summation in the last line. We now apply the third condition to conclude that the last line is zero.

\clearpage

\end{document}